\documentclass[12pt]{article}
\usepackage[utf8]{inputenc}

\usepackage{setspace}

\usepackage[letterpaper]{geometry}
\usepackage{apacite}
\usepackage{natbib} % reference manager

\usepackage{hyperref}
\usepackage[affil-it]{authblk}

\usepackage{amsthm,amsmath,amsfonts,amssymb}
\usepackage{graphicx}% uncomment this for including figures
\usepackage{bm} % for bold in math env
\usepackage{threeparttable}% for tablenotes
\usepackage{float}
\usepackage{caption}
\usepackage{cancel} % to strike out formulas
\usepackage{makecell} % to write on multiple lines in a table
\usepackage{comment}
\usepackage{xcolor}

\DeclareCaptionLabelSeparator*{spaced}{\\[2ex]}
\theoremstyle{remark}
\newtheorem{assumption}{Assumption}
\usepackage{todonotes}
\usepackage{xcolor,cancel} % to cancel math formulas

\def\independent{\protect\mathpalette{\protect\independenT}{\perp}}
\def\independenT#1#2{\mathrel{\rlap{$#1#2$}\mkern2mu{#1#2}}}

\begin{document}
	
	\title{\textbf{Exploiting Network Information to Disentangle Spillover Effects in a Field Experiment on Teens' Museum Attendance}}
	
	\author[1]{Silvia Noirjean}
	\author[2]{Marco Mariani}
	\author[1,3]{Alessandra Mattei}
	\author[1,3]{\\Fabrizia Mealli}
	
	\affil[1]{\small{Department of Statistics, Computer Science, Applications ``G. Parenti", University of Florence}}
	\affil[2]{\small{IRPET - Regional Institute for Economic Planning of Tuscany}}
	\affil[3]{\small{Florence Center for Data Science}}
	
	\date{} % Comment this line to show today's date
	
\maketitle

\doublespacing
\begin{abstract}
%	\noindent \textit{\textcolor{red}{A key element in the education of youths is their sensitization to historical and artistic heritage. We analyze a field experiment conducted in Florence (Italy) to assess how appropriate incentives assigned to high-school classes may induce teens to visit museums in their free time. The presence of non-compliance and spillover effects makes an impact evaluation of this clustered encouragement design particularly challenging. We propose to blend the principal stratification framework and causal mediation, by defining sub-populations of units according to their compliance behavior and using the information on their friendship networks as mediator. We formally define principal natural direct and indirect effects and principal controlled direct and spillover effects, and use them to disentangle spillovers from other causal channels. We adopt a Bayesian approach for inference.}}
A key element in the education of youths is their sensitization to historical and artistic heritage. We analyze a field experiment conducted in Florence (Italy) to assess how appropriate incentives assigned to high-school classes may induce teens to visit museums in their free time. Non-compliance and spillover effects make the impact evaluation of this clustered encouragement design challenging. We propose to blend principal stratification and causal mediation, by defining sub-populations of units according to their compliance behavior and using the information on their friendship networks as mediator. We formally define principal natural direct and indirect effects and principal controlled direct and spillover effects, and use them to disentangle spillovers from other causal channels. We adopt a Bayesian approach for inference.
  
\end{abstract}
\noindent Keywords: \textit{Bayesian inference; causal inference with network interference; clustered encouragement design; mediation analysis; principal stratification}

%\noindent Keywords: \textit{Bayesian inference; causal inference with interference; \textcolor{red}{\cancel{clustered encouragement design;}} mediation analysis; networks; principal stratification}

% to avoid numbering the first page
\thispagestyle{empty} 
\clearpage
\pagenumbering{arabic} 

\newpage
\section{Introduction}
\label{intro}

Over the last years, there has been an increasing interest in causal studies where units are part of networks (e.g., groups of friends). In this type of studies, social or physical interactions among units are a major concern, and thus, the no-interference assumption, under which  the outcome of each unit cannot depend on the treatment assigned to others \citep{Cox:1958}, is no longer plausible. 

The literature on causal inference in the presence of networks has developed considerably in recent years \citep[e.g.][]{Goldsmith:2013,VanderWeele_An:2013,Kang:2016,Ogburn:2017a,Forastiere:2018,Forastiere:2021}.  %\citep[e.g.][]{Goldsmith:2013,VanderWeele_An:2013,Kang:2016,Ogburn:2017a,Ogburn:2017b,Forastiere:2018,Forastiere:2021}. 
The intuition characterizing this literature is that interference between units can be conveyed by the specific social links that tie these units together into a network, irrespective of treatment assignment. With this in mind, it is as if each unit was exposed to multiple treatments: its own treatment and the treatments of the other units in its network. The latter may give rise to spillover effects \citep{DelPrete:2019}. 

Interference naturally arises in clustered encouragement designs (CEDs), where encouragement is randomized at the level of a cluster (e.g., class) of subjects (e.g., students) but compliance with the assigned encouragement is often at the individual level. 
CEDs are widely adopted, especially when randomization of the receipt of treatment is not feasible for ethical or practical reasons and the assignment of encouragement at the level of clusters appears preferable for administrative convenience, cost reduction, ethical issues and for improving treatment adherence. The literature abounds of studies designed as CEDs with individual noncompliance   \cite[e.g.,][]{Frangakis:2002, Morris:2004, Forastiere:2016}.
% \cite[e.g.][]{McDonald:1992, Hirano:2000, Frangakis:2002, Morris:2004}.

Clusters define a special network, where sets of units belonging to the same cluster are likely to share neighbors, but have no links with units belonging to different clusters.
Therefore, in CEDs	social interactions naturally occur among units belonging to the same group, and we can reasonably expect that among subjects who interact with one another, the treatment actually received by one subject may affect other subjects' outcomes, generating interference or spillover effects \citep[]{Sobel:2006, Hong:Raudenbush:2006, Hudgens:Halloran:2008}.
The presence of interference gives rise to intricate causal mechanisms, and disentangling them may provide precious information for researchers and policymakers to understand how an intervention works and how it can be improved.

A CED with individual noncompliance is the study design underlying our motivating study on teens' museum attendance. It is a field experiment based on an encouragement design clustered at the level of school classes, conducted in Florence (Italy) and designed to asses how appropriate incentives may prompt teens to visit museums in their free time. Throughout the text we refer to this study as the Florentine study. A unique feature of this study is the availability of information about the friendship ties between the teens belonging to the same class, which makes the study a \textit{CED with a hierarchical network structure}, with a high-level network defined by the clustered structure and a low-level network defined by the friendship structures within each cluster.
The study was previously analyzed by \cite{Lattarulo:2017} and \cite{Forastiere:2019}, but they neglect information on friendship ties, and thus, provide only a partial picture of the causal mechanisms by which the designed incentives affect teens' museums attendance.

We make important substantive contributions by adopting an innovative approach to the analysis of CEDs with individual noncompliance under network interference. Teens' museums attendance is a relevant topic because, while teens are likely to visit museums during school or family trips, museums are seldom their choice when it is up to them to choose where to spend their free time \cite[e.g.][]{Hughes:2019}. %\cite[][]{Gray:1998, Hughes:2019}. 
In addition to granting free or low-price entrance to youths \cite[][]{Cellini:2018}, many museums are attempting to portray an image of educational and entertaining institutions and devising strategies to attract teens and young adults towards a tailored visit experience \cite[][]{Lattarulo:2017,Manna:2018}. The hope is that such experience can raise further engagement with museums and cultural heritage in general \cite[][]{Kisida:2014}.
In the Florentine experiment, classes of students are randomized to three forms of encouragement aimed to prompt students to visit Palazzo Vecchio, one of the most famous museums in the city. Students belonging to a first group of classes receive a flyer with basic information about Palazzo Vecchio, including its opening hours. Students in a second group of classes receive both the flyer and a short presentation conducted by an art expert. The goal of the presentation is to enhance students' curiosity about museum visits, in general, and Palazzo Vecchio, in particular. Students in a third group of classes, in addition to the flyer and the presentation, receive a non-financial reward in the form of extra-credit points towards their school grade. 
The treatment is the actual visit to Palazzo Vecchio, which can be performed in the two months after the encouragement. The outcome of interest is the number of museum visits performed in the follow-up period, that is, between two and eight months from the encouragement.

The stronger the encouragement, the higher the propensity of a student to visit Palazzo Vecchio is expected to be. However, students' compliance with the cluster-level encouragement is imperfect, meaning that there are both students in classes assigned to the flyer who visit Palazzo Vecchio as well as students in classes assigned to the reward who do not visit Palazzo Vecchio.
We deal with the presence of noncompliance using the principal stratification (PS) framework \cite[][]{Frangakis:Rubin:2002}: we define stratum-specific causal effects, named principal causal effects, which are effects for specific latent subpopulations, defined by the joint potential compliance statuses under the three encouragement conditions.
The role of PS in encouragement designs to draw inference on intention-to-treat effects within principal strata is uncontroversial 
\cite[e.g.,][]{Angrist:1996, Imbens:Rubin:1997, Hirano:2000, Frangakis:2002}. %\cite[e.g.,][]{Imbens:Angrist:1994, Angrist:1996, Imbens:Rubin:1997, Frangakis:2002}.

How to deal with network interference in CEDs is still an open research issue.  In principle, a number of causal pathways may take place within CEDs. The first path is the effect of the encouragement on a unit's outcome passing through that unit's treatment uptake. The second path is through spillover effects, i.e., changes in a unit's outcome originating from the treatment uptake by other units, which naturally arise among friends.
The third path is the direct effect of the encouragement passing neither through the individual treatment uptake nor through spillovers, which ends up pooling further causal channels.

Two recent studies by \cite{Forastiere:2016} and \cite{Forastiere:2019} provide insightful contributions on the causal mechanisms in action in CEDs.
\cite{Forastiere:2016} propose to use the principal stratification framework to disentangle distinct causal mechanisms in CEDs, viewing the treatment variable as a mediating variable and defining causal effects conditional on principal stratum membership based on hypothetical interventions on the treatment uptake.
\cite{Forastiere:2019} show how  intention-to-treat effects within principal strata can be interpreted to get valuable information on different causal mechanisms for different types of units.
Both \cite{Forastiere:2016} and \cite{Forastiere:2019} find evidence  that  encouragement and treatment effects in their CEDs are blended with spillovers, and highlight that disentangling these effects may be uneasy. We contribute to this literature providing an innovative method for untying  encouragement, treatment and spillovers in CEDs with  a hierarchical network structure, which capitalize on the distinguish structure of our data: the availability of  within cluster network data.  
Specifically, our dataset includes information on the friendship network within the class of each student participating in the study. We use this information to construct a new variable defined as the proportion of friends who actually take the treatment, that is, visit Palazzo Vecchio.
Our key insight is to use this variable as a mediator and conduct  principal stratum mediation analysis to disentangle distinct causal mechanisms. We define principal stratum natural direct and indirect effects and principal stratum controlled direct effects, which are natural direct and indirect effects and controlled direct effects \cite[][]{RobinsGreenland:1992, Pearl:2001} for subpopulations of units defined by the principal stratification with respect to the individual receipt of treatment.
We also introduce new causal estimands, which we refer to as principal controlled spillover effects, defined as comparisons within principal strata of hypothetical potential outcomes for the response variable under a fixed level of the encouragement and different values of the mediator. We interpret principal natural indirect effects and principal controlled spillover effects as spillover effects, which may be heterogeneous across different types of units defined by the compliance statuses. 
Principal natural and controlled direct effects, instead, can be viewed as pure effect of the encouragement or as a blend of the encouragement and the treatment/experience effects, depending on the principal stratum and the encouragement conditions being compared.

Inference on the causal estimands of interest is conducted under latent ignorability assumptions of the mediator using a Bayesian approach.
We introduce two latent ignorability assumptions of the mediator: a latent ignorability assumption of the observed mediator, implying   that the observed  proportion of friends visiting Palazzo Vecchio are independent of potential outcomes for the number of museum visits conditional on principal stratum membership given the observed value of the encouragement and individual level and network level pretreatment covariates; and a latent ignorability assumption of the potential mediator, implying that potential outcomes for the  proportion of friends visiting Palazzo Vecchio are independent of potential outcomes for the number of museum visits conditional on principal stratum membership given observed individual level and network level pretreatment covariates.	
Under latent ignorability assumptions, ignorability of the mediator only holds after conditioning on principal stratum membership, which is a latent variable, only partially observed for each unit.
%If principal stratum membership were fully observed for each unit, results from the   literature on causal mediation analysis could be used to identify and estimate  principal stratum natural direct and indirect effects and principal stratum controlled direct and spillover effects. 
The latent nature of the principal strata complicates inference. We deal with inferential issues using a Bayesian approach, which does not require full identification \cite[e.g.][]{Gustafson:2010}. 
%In Bayesian inference focus is on deriving the posterior distribution of the parameters of interest, by updating a prior distribution via a likelihood, irrespective of whether the parameters are fully or partially identified; the posterior distribution is always proper if the prior distribution is proper \cite[e.g.][]{Gustafson:2010}. 
We specify flexible parametric models for the principal stratum membership given individual- and network-level covariates and for the potential outcomes conditional on principal stratum membership and individual- and network-level covariates. 

Our work provide a cutting-edge innovative approach to the analysis of CEDs, which smartly blend concepts and tools from principal stratification, mediation analysis and causal inference with network data.

The article is organized as follows. Section 2 introduces the experimental design and the data. Section 3 describes the methodology. 
%In particular, it starts by setting the necessary notation. Later, it presents the main components of our proposal, that are principal stratification, mediation analysis and network information, and it explains how to blend them. Then, it ends by reporting the models adopted for the estimation of the causal effects of interest. 
Section 4 reports the results of the analysis and Section 5 concludes.

\section{Experimental Design and Data}
\label{sec:experimentaldesign}

The clustered encouragement field experiment we re-analyze was run in Florence, Italy, at three different points in time during 2014. It involved $J=15$ classes, indexed by $j=1,...,15$, from 3 different high schools in the city. Let $N_j$ be the number of students in class $j$, $j=1,...,15$, and let $N=\sum_{j=1}^{J}N_j$ be the total number of students participating in the study: $N=266$. Finally, let $C_{ij}$ be an indicator for student $i$'s class membership: $C_{ij}=1$ if student $i$ is in class $j$ and zero otherwise, $i=1, \ldots, N_j$ and $j=1, \ldots, J$. All high schools were of the same type and all students in the classes participating in the study were aged 17-18. Each of the 15 classes was randomly assigned to one of three incremental encouragements, in groups of five. All students, irrespective of the encouragement their class was assigned to, were offered the opportunity to visit individually the Florentine Museum of Palazzo Vecchio outside of school hours. 

Under the weakest encouragement, which we name \textit{flyer}, students received a flyer containing basic information about Palazzo Vecchio (opening hours and a brief description of the museum), and a short text written by the experimenters, briefly stating the importance of museum attendance. Under the intermediate encouragement, named \textit{presentation}, in addition to the flyer and text, students received a short presentation conducted by an art expert from the museum: the aim of the presentation was to enhance students' curiosity about museum visits, in general, and Palazzo Vecchio, in particular, and to portray museum attendance as an intriguing and entertaining experience. Under the strongest encouragement, which we name \textit{reward}, in addition to the flyer, text and presentation, students were promised a non-financial reward in the form of extra-credit points towards their final school grade if they actually visited Palazzo Vecchio within two months after assignment.
Let $Z_j$ denote the encouragement which class $j$ is assigned to: $Z_j$=1 if the class is assigned to the \textit{flyer}, $Z_j$=2 if it is assigned to the \textit{presentation} and $Z_j$=3 if it is assigned to the \textit{reward}.  

The field experiment was implemented through three visits to the classes, whose timing and contents are described in what follows. At time $t = 1$, all students were surveyed about their background characteristics and interviewed about their own friendship network within the class.
In particular, they were asked to mention all the classmates they consider their friends. Then, classes were randomly assigned to the three encouragement levels described above. All students were offered a free visit to Palazzo Vecchio to be made within two months.

At time $t = 2$, two months after the assignment, information about whether students had visited or not Palazzo Vecchio was collected. Let $M_{ij}$ be a binary variable, equal to 1 if student $i$ in class $j$ visits Palazzo Vecchio in the two months after the assignment, and zero otherwise. The variable $M_{ij}$ is the treatment of interest.

A follow-up period of 6 months, starting from time $t = 2$ was considered. At the end of the study, that is, at time $t = 3$ (6 months after $t = 2$), information on the number of visits each student made to other museums in the follow-up period was collected. 
Let $Y_{ij}$  be a count variable  indicating the number of  museum visits made by student $i$ in class $j$ during the follow-up period: $Y_{ij}$ is the outcome of primary interest. See  \cite{Lattarulo:2017} and \cite{Forastiere:2019}
for a more detailed description of the three forms of encouragement and a comprehensive analysis of the substantive results of the field experiment.

\cite{Lattarulo:2017} use classes as units of analysis and resort to randomization inference techniques to conduct an intention-to-treat analysis of the differential effects of the three encouragements on classroom-level museum attendance. \cite{Forastiere:2019}, instead, use students as units of analysis considering the students' choice to visit  Palazzo Vecchio as an endogenous treatment. Under the partial interference assumption \cite[e.g.,][]{Sobel:2006}, they use PS to investigate the heterogeneity of causal effects across different latent subgroups of units. Although they find that, for all these subgroups, encouragement and/or treatment effects are mixed with spillovers, disentangling these effects is beyond the aim of their study. Our methodological contribution fills this gap using information on individual friendship networks, which is not exploited in \cite{Forastiere:2019}.

Let us denote students' friendship network as a pair ($\mathcal{N}, \mathcal{E}$), where $\mathcal{N}$ is the set of nodes (subjects) and $\mathcal{E}=\{E_{ik}, i, k=1, \ldots, N_j; j=1, \ldots, J\}$ is the set of edges (friendship ties) that links the subjects; in particular,
$E_{ik} = E_{ki} = 1$ if subject $i$ declared subject $k$ as a friend and/or viceversa, and $E_{ik} = E_{ki} = 0$ otherwise. It is worth noting that, by design, $E_{ik} = E_{ki} = 0$ for each pair of students $i$ and $k$ belonging to two different classes. Therefore we are considering a binary and undirected network: ``binary'' because we assign an equal weight to all the friendships, and ``undirected'' because an influence between subject $i$ and subject $k$ is plausible if at least one of them declared the other as a friend. 
For each student $i$ in class $j$, let $\mathcal{N}_{ij}$ denote the neighborhood of student $i$, comprising student $i$'s friends in class $j$.

For each student $i$ in class $j$, $i=1\ldots, N_j$, $j=1,\ldots, J$, we define a new variable $S_{{\mathcal{N}}_{ij}}$ as the proportion of student $i$'s friends who visit Palazzo Vecchio.  $S_{{\mathcal{N}}_{ij}}$ is a summary of $M_{i'j}$ for all $i' \neq i \in \mathcal{N}_{ij}$:

{\centering
	$ \displaystyle
	S_{{\mathcal{N}}_{ij}}= \dfrac{1}{|\mathcal{N}_{ij}|}\sum_{i' \neq i \in \mathcal{N}_{ij}} M_{i'j}
	$ 
\par}
\noindent where $|\mathcal{N}_{ij}|$ is the number of student $i$'s friends, that is, the degree of student $i$.

For each  student $i$ in class $j$, $i=1\ldots, N_j$, $j=1,\ldots, J$, we observe a vector of $K$ covariates,  $\bm{X}_{ij}$. This vector is composed by two sub-vectors: $\bm{X}_{ind_{ij}}$ and  $\bm{X}_{\mathcal{N}_{ij}}$, which include individual and neighborhood-level covariates, respectively.
The sub-vector $\bm{X}_{\mathcal{N}_{ij}}$ includes different types of neighborhood-level covariates: some of them are a synthesis of individual-level covariates in the neighborhood, such as, e.g., the proportion of male friends or the average number of museum visits made by friends prior to the experiment; others represent the structure of the neighborhood, such as the degree (number of ties) of a unit or the average degree of a unit's friends.

\subsection{Descriptive statistics}
Table \ref{tab:covariates} reports some descriptive statistics for the sample of the 266 students, their friends and the structure of their individual friendship network, grouped by class assignment, $Z_j$. 
\begin{table}[!ht]
	\centering
	%    \caption{Descriptive statistics, prior and post-encouragement (proportions or means).}
	\caption{The museum study: descriptive statistics of the individual and friends' level background covariates (proportions or means)}
	\label{tab:covariates}
	\resizebox{\columnwidth}{!}{%
		\begin{tabular}{lcccc}
			\hline\noalign{\smallskip}
			& \textit{Flyer} & \textit{Presentation} & \textit{Reward} & \textit{Overall} \\
			\noalign{\smallskip}\hline\noalign{\smallskip}
			\textit{INDIVIDUAL BACKGROUND CHARACTERISTICS}\\ [+0.1cm]
			Male (1/0) & 0.19 & 0.29 & 0.53 & 0.34 \\ [+0.1cm]
			Already visited Palazzo Vecchio (1/0) & 0.66 & 0.75 & 0.72 & 0.71 \\ [+0.1cm]
			No. of museums visited previous year & 3.31 & 4.75 & 3.54 & 3.86 \\ [+0.1cm]
			GPA (0-10)\textsuperscript{a} & 6.73 & 6.82 & 6.98 & 6.84 \\ [+0.1cm]
			Interest in sciences (1-3)\textsuperscript{b} & 2.12 & 1.74 & 2.23 & 2.03 \\ [+0.1cm]
			Parental education (1-5)\textsuperscript{c} & 3.26 & 3.54 & 3.56 & 3.45 \\
			\noalign{\smallskip}\hline\noalign{\smallskip}
			\textit{FRIENDS' BACKGROUND CHARACTERISTICS}\\ [+0.1cm]
			Male (1/0) & 0.18 & 0.28 & 0.49 & 0.32 \\ [+0.1cm]
			Already visited Palazzo Vecchio (1/0) & 0.62 & 0.74 & 0.69 & 0.68 \\ [+0.1cm]
			No. of museums visited previous year & 3.07 & 4.59 & 3.20 & 3.61 \\ [+0.1cm]
			GPA (0-10)\textsuperscript{a} & 6.66 & 6.78 & 6.61 & 6.68 \\ [+0.1cm]
			Interest in sciences (1-3)\textsuperscript{b} & 2.09 & 1.68 & 2.06 & 1.95 \\ [+0.1cm]
			Parental education (1-5)\textsuperscript{c} & 3.23 & 3.59 & 3.44 & 3.42\\
			\noalign{\smallskip}\hline\noalign{\smallskip}
			\textit{STRUCTURE OF FRIENDSHIP NETWORKS}\\ [+0.1cm]
			No. of friends (individual degree) & 4.94 & 4.00 & 3.07 & 4.00 \\[+0.1cm] 
			No. of friends of friends (friends' degree) & 5.62 & 4.81 & 3.43 & 4.61  \\ [+0.1cm]
			\noalign{\smallskip}\hline\noalign{\smallskip}
			No. of classes & 5 & 5 & 5 & 15 \\ [+0.1cm]
			No. of students per class & 17.8 & 17.4 & 18 & 17.7 \\[+0.1cm]
			No. of students & 89 & 87 & 90 & 266 \\
			\noalign{\smallskip}\hline
		\end{tabular}
	}
	\begin{flushleft}
		\textit{Note}. Estimates are based on sampling-theory estimation of a ratio for cluster sampling \citep[pag. 30]{Cochran:1963}.
		\begin{tablenotes}[flushleft]
			\item[] \textsuperscript{a} Grade Point Average (GPA) is continous on a 1-10 point scale.
			\item[] \textsuperscript{b} Levels of interest in sciences: 1 if the interest is only in humanities, 2 if it is in both humanities and sciences, 3 if it is only in sciences.
			\item[] \textsuperscript{c} Levels of parental education: 1 if both parents completed only compulsory education, 2 if only one of them completed high school, 3 if both completed high school, 4 if only one of them completed university, 5 if both parents completed university.
		\end{tablenotes}
	\end{flushleft}
	
\end{table}
As it is often the case with small-scale field experiments, background characteristics are not always well balanced across encouragement groups. Imbalances between groups of students in classes assigned to the alternative encouragement levels arise in almost all the individual-level and network(friends)-level background characteristics. In particular, number of museums visited the previous year and proportion of males differ remarkably  across students in classes exposed to  alternative encouragement levels.
Students assigned to presentation visited more museums, on average, than those assigned to the other encouragements; not surprisingly, they are also more interested in the humanities. Furthermore, the proportion of males is much higher among students assigned to the reward. Imbalances also arise in the structure of friendship networks. Specifically, the average number of ties (i.e., friends) is about 5, 4 and 3 in the flyer, presentation and reward  group, respectively. These differences result in a different number of friends of friends across encouragement groups, too: students assigned to the reward tend to have less friends -- and less friends of friends -- than those assigned to the presentation and, to a greater extent, than those assigned to the flyer.

\begin{figure}[ht]
	\centering
	\caption{Outcome bar charts}
	\includegraphics[width=0.9\columnwidth]{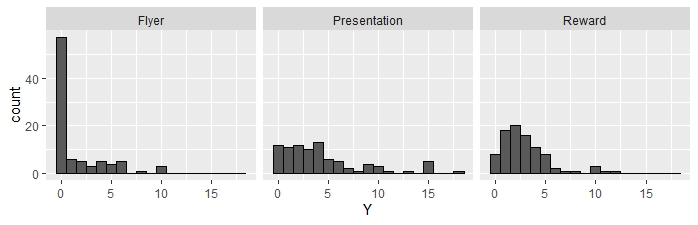}
	\label{fig:Y}
\end{figure}

Table \ref{tab:outcomes} reports descriptive statistics regarding treatment uptake (Palazzo Vecchio visit) and final outcome (number of museum visits during the follow-up), again at both individual/student and network/student's friends level. In addition, Figure \ref{fig:Y} shows the bar charts of the final outcome by encouragement status.
\begin{table}[ht]
	\centering
	\caption{The museum study: descriptive statistics of the treatment and the outcome variables}
	\label{tab:outcomes}
	\resizebox{\columnwidth}{!}{%
		\begin{tabular}{lcccc}
			\hline\noalign{\smallskip}
			& \textit{Flyer} & \textit{Presentation} & \textit{Reward} & \textit{Overall} \\
			\noalign{\smallskip}\hline\noalign{\smallskip}
			\textit{INDIVIDUAL TREATMENT  AND OUTCOME}\\ [+0.1cm]
			No. of students performing the proposed visit & 3 & 10 & 40 & 53 \\ [+0.1cm]
			Share of students performing the proposed visit & 0.03 & 0.11 & 0.44 & 0.20 \\ [+0.1cm]
			Count of museum visits during the follow-up & 1.49 & 4.39 & 3.00 & 2.95 \\ [+0.1cm]
			Students visiting at least one museum during the follow-up & 0.36 & 0.86 & 0.91 & 0.71 \\
			\noalign{\smallskip}\hline\noalign{\smallskip}
			\textit{FRIENDS' TREATMENT  AND OUTCOME}\\ [+0.1cm]
			Share of friends performing the proposed visit & 0.03 & 0.10 & 0.41 & 0.18  \\ [+0.1cm]
			Count of museum visits during the follow-up & 1.57 & 4.02 & 2.86 & 2.80  \\ [+0.1cm]
			Friends visiting at least one museum during the follow-up & 0.56 & 0.98 & 0.92 & 0.82  \\
			\noalign{\smallskip}\hline
		\end{tabular}
	}
\end{table}
The percentage of students who visit Palazzo Vecchio after receiving the flyer is only 3\%. Among students who also attend the presentation made by the art expert, this percentage increases  to 11\%. A sharper increase is observed when the reward is also promised in the event of a visit to Palazzo Vecchio: in this case, in fact, 44\% of students performed the proposed visit. 
However, it is worth noting that the interest of the field experiment is not on investigating which incentive could better motivate students to perform a particular museum visit once, but on assessing which incentive is more effective to prompt teens' museums attendance in general. Simple comparisons of the outcomes by encouragement status suggest that the flyer is insufficient to motivate students to visit art museums in the follow-up period, just as it is insufficient to motivate students to visit Palazzo Vecchio in the first place. As we can see in Figure \ref{fig:Y}, the empirical distributions of the number of museum visits during the follow up by encouragement are left-skewed and have a mass at zero. The proportion of students  who do not visit any museum is especially high under the flyer encouragement. The number of museum visits has a similar distribution under the presentation and reward encouragements, which seem to be more effective than the flyer encouragement.  The average number of total museum visits is
   relatively high under  both the presentation and the reward (4.39 and 3.00, respectively). %Although the average number of total museum visits is higher among students assigned to the presentation than among students assigned to the reward (4.39 versus 3.00), it is relatively high under both encouragement levels. 
Therefore, there is some evidence that these two types of encouragement may have at least partially achieved the purpose of motivating students to visit art museums, regardless of whether the triggered mechanism is ascribable to the encouragement itself, to the treatment or to spillovers from friends. 
 
\section{Methodology}
\subsection{Potential outcomes}
Let $\bm{z}$ denote a $J-$dimensional vector of the encouragement assignments for the whole sample of classes with $j$th element equal to $z_j$: $\bm{z} \in \{1,2,3\}^J$. For each student $i$ in class $j$, $i=1\ldots, N_j$, $j=1,\ldots, J$, let $M_{ij}(\bm{z})$ denote the potential indicator for visiting Palazzo Vecchio given the encouragement vector $\bm{z}$; also let $Y_{ij}(\bm{z})$ denote the potential outcome of main interest, that is, the number of museums visited in the follow-up period  given the encouragement vector $\bm{z}$. 
Because $S_{{\mathcal{N}}_{ij}}$ is function of $M_{i'j}$,  $i' \neq i \in \mathcal{N}_{ij}$, it is a post-encouragement variable, and thus, we introduce potential outcomes for it, too:

{\centering
	$ \displaystyle
	S_{{\mathcal{N}}_{ij}}(\bm{z}) = \dfrac{1}{|\mathcal{N}_{ij}|}\sum_{i' \neq i \in \mathcal{N}_{ij}} M_{i'j}(\bm{z}).
	$
	\par}

We assume that a cluster-level Stable Unit Value Assumption \citep[SUTVA,][]{Rubin:1980} holds for the  potential outcomes $M_{ij}(\bm{z})$, $Y_{ij}(\bm{z})$, and $S_{{\mathcal{N}}_{ij}}(\bm{z})$:

\begin{assumption}
	\label{ass:sutva}(SUTVA for $M_{ij}(\bm{z})$, $Y_{ij}(\bm{z})$, and $S_{ij}(\bm{z})$) 
	
	\noindent \textit{No hidden variation of encouragement}: For all $\bm{z}, \bm{z}'$: $\bm{z}=\bm{z}'$, $M_{ij}(\bm{z})=M_{ij}(\bm{z}')$,  $Y_{ij}(\bm{z})=Y_{ij}(\bm{z}')$, $S_{{\mathcal{N}}_{ij}}(\bm{z})=S_{{\mathcal{N}}_{ij}}(\bm{z}')$.
	
	\noindent \textit{Cluster-level no-interference}:  For all $\bm{z}, \bm{z}'$: $z_j = z_j'$, $M_{ij}(\bm{z})=M_{ij}(\bm{z}')$,  $Y_{ij}(\bm{z})=Y_{ij}(\bm{z}')$, $S_{{\mathcal{N}}_{ij}}(\bm{z})=S_{{\mathcal{N}}_{ij}}(\bm{z}')$.
\end{assumption}

Assumption \ref{ass:sutva} allows to write $M_{ij}(\bm{z})$,  $Y_{ij}(\bm{z})$ and  $S_{{\mathcal{N}}_{ij}}(\bm{z})$  as
$M_{ij}(z_j)$,  $Y_{ij}(z_j)$ and  $S_{{\mathcal{N}}_{ij}}(z_j) = |\mathcal{N}_{ij}|^{-1}\sum_{i'\neq i \in \mathcal{N}_{ij}} M_{i'j}(z_j)$.
Assumption \ref{ass:sutva} implies that the outcomes of a given student does not vary with the encouragement assigned to students of other classes. Under SUTVA, $M_{ij} = M_{ij}(Z_j)$, $S_{ij} = S_{ij}(Z_j)$, and $Y_{ij} = Y_{ij}(Z_j)$.

Our key insight is that spillover effects may pass through the variable $S_{{\mathcal{N}}_{ij}}$.
We formalize this intuition by viewing  it as a mediator.
For $j=1, \ldots, J$, let $\bm{S}_{j}$ and $\bm{S}_{j}(z_j)$ denote  $N_j-$dimensional vectors with $i$th element equal to $S_{{\mathcal{N}}_{ij}}$ and $S_{{\mathcal{N}}_{ij}}(z_j)$, respectively,  $i=1\ldots, N_j$. For each student $i$, we assume the existence of potential outcomes of the form $Y_{ij}(z_j, \bm{s}_j)$ and $Y_{ij}(z_j, \bm{S}_j(z_j'))$: $Y_{ij}(z_j, \bm{s}_j)$ would be the value of the outcome $Y$ if, possibly contrary to fact, the encouragement were set to the level $z_j$ and the mediator vector, $\bm{S}_{j}$, were set to a specific prefixed value, $\bm{s}_j$; and $Y_{ij}(z_j, \bm{S}_j(z_j'))$ would be the value of the outcome $Y$ if, possibly contrary to fact, the encouragement were set to the level  $z_j$  and the mediator vector, $\bm{S}_{j}$, were set to the value it would have taken if  the encouragement  had been set to an alternative level, $z_j'$.

We believe to be plausible that a cluster version of the \textit{Stable Unit Treatment on Neighborhood Value Assumption} (SUTNVA) introduced by \cite{Forastiere:2021} holds for potential outcomes of the form 
$Y_{ij}(z_j, \bm{s}_j)$ and $Y_{ij}(z_j, \bm{S}_j(z_j'))$. Formally, 

%\begin{assumption} \label{ass:psutnva}(Cluster SUTNVA) 
%	
%	\noindent \textit{No hidden variation of \sout{encouragements and} mediators}: 
%	
%	\noindent For all $\bm{Z}, \bm{Z}'$: $\bm{Z}=\bm{Z}'$ and for all  $\bm{s},  \bm{s}'$:
%	$\bm{s}= \bm{s}'$
%	$$
%	Y_{ij}(\bm{Z}, \bm{s}) = Y_{ij}(\bm{Z}', \bm{s}') $$
%	
%	\noindent For all $\bm{Z}, \bm{Z}''$: $\bm{Z}=\bm{Z}''$ and for all $\{\bm{S}_j(Z_j')\}_{j=1}^J,  \{\bm{S}'_j(Z_j')\}_{j=1}^J$:
%	$\{\bm{S}_j(Z_j')\}_{j=1}^J= \{\bm{S}'_j(Z_j')\}_{j=1}^J$
%	$$  
%	Y_{ij}(\bm{Z}, \{\bm{S}_j(Z_j')\}_{j=1}^J) = Y_{ij}(\bm{Z}'', \{\bm{S}'_j(Z_j')\}_{j=1}^J)
%	$$   	
%	
%	\noindent \textit{Cluster Neighborhood Interference}: 
%	
%	\noindent If $Z_j = Z_j'$ and  $s_{ij}=s'_{ij}$,  then
%	$$
%	Y_{ij}(\bm{Z}, \bm{s}) = Y_{ij}(\bm{Z}', \bm{s}')$$
%	
%	\noindent If $Z_j = Z_j''$ and  $S_{{\mathcal{N}}_{ij}}(Z_j')=S'_{{\mathcal{N}}_{ij}}(Z_j')$, then
%	$$
%	Y_{ij}(\bm{Z}, \{\bm{S}_j(Z_j')\}_{j=1}^J) = Y_{ij}(\bm{Z}'', \{\bm{S}_j(Z_j')\}_{j=1}^J)
%	$$
%\end{assumption}  

\begin{assumption} \label{ass:psutnva}(Cluster SUTNVA) 
	
	\noindent \textit{No hidden variation of  mediators}: 
	
	\noindent For all   $\bm{s}_j,  \bm{s}_j'$: $\bm{s}_j=\bm{s}_j'$, $Y_{ij}(z_j, \bm{s}_j) = Y_{ij}(z_j, \bm{s}'_j)$.

	\noindent For all  $\bm{S}_j(z'_j),\bm{S}_j(z''_j)$: $\bm{S}_j(z'_j)=\bm{S}_j(z''_j)$, $Y_{ij}(z_j, \bm{S}_j(z'_j)) = Y_{ij}(z_j, \bm{S}_j(z''_j))$.
	
	\noindent \textit{Cluster Neighborhood Interference}: 
	
	\noindent For all $\bm{s}_j,  \bm{s}_j'$: $\bm{s}_{\mathcal{N}_{ij}}= \bm{s}_{\mathcal{N}_{ij}}'$, $Y_{ij}(z_j, \bm{s}_j) = Y_{ij}(z_j, \bm{s}'_j)$.
	
	\noindent For all  $\bm{S}_j(z'_j),\bm{S}_j(z''_j)$:
	$ S_{\mathcal{N}_{ij}}(z_j')=S_{\mathcal{N}_{ij}}(z_j'')$, $Y_{ij}(z_j, \bm{S}_j(z'_j)) = Y_{ij}(z_j, \bm{S}_j(z''_j))$.
	
\end{assumption}  

Cluster SUTNVA implies that  potential outcomes  of the form $Y_{ij}(z_j, \bm{s}_j)$ and $Y_{ij}(z_j, \bm{S}_j(z_j'))$ for student $i$ in class $j$ depend on proportion of student $i$'s friends who visit Palazzo Vecchio, but  do not vary with the proportion of friends of  other classmate students who visit Palazzo Vecchio.
Under cluster SUTNVA (Assumption~\ref{ass:psutnva}), we can write $Y_{ij}(z_j,  \bm{s}_j)$ and 
$Y_{ij}(z_j,\bm{S}_j(z_j'))$ as  $Y_{ij}(z_j, s_{\mathcal{N}_{ij}})$ and
$Y_{ij}(z_j, S_{{\mathcal{N}}_{ij}}(z_j'))$, respectively. Note that, under Assumptions \ref{ass:sutva} and \ref{ass:psutnva},  $Y_{ij}=Y_{ij}(Z_j)=Y_{ij}(Z_j, S_{{\mathcal{N}}_{ij}}(Z_j))$.

It is worth noting that  we have to know the network that links the units to exploit Assumptions~\ref{ass:sutva} and \ref{ass:psutnva}, which are based on the idea that interference can take place between immediate neighbors. Moreover we need to assume that this network is fixed and that the links between nodes are correctly measured. In our setting this assumption means that any change in the vector of assignment $\bm{z}$ and in $\{\bm{S}_j(z_j)\}_{j=1}^J$ does not modify the network that links the subjects.

In the causal inference literature, potential outcomes of the form $Y_{ij}(z_j, s_{\mathcal{N}_{ij}})$ and $Y_{ij}(z_j, S_{{\mathcal{N}}_{ij}}(z_j'))$ are also called \textit{a priori} counterfactual potential outcomes \citep[e.g.,][]{Frangakis:Rubin:2002}, because some of them can never be observed for some units. For instance, \textbf{$Y_{ij}(z_j,S_{\mathcal{N}_{ij}}(z'_{j}))$} cannot be observed for any units with  \textbf{$S_{\mathcal{N}_{ij}}(z'_j)\neq S_{\mathcal{N}_{ij}}(z_j)$}.
In the sequel, we denote by $M_{ij}(z)$, 	$S_{{\mathcal{N}}_{ij}}(z)$ and $Y_{ij}(z)$ potential outcomes for $M_{ij}$, 	$S_{{\mathcal{N}}_{ij}}$ and $Y_{ij}$ for a student $i$ in class $j$  under class encouragement $z$. Similarly, we use the notation $Y_{ij}(z, s_{\mathcal{N}_{ij}})$ and $Y_{ij}(z, S_{{\mathcal{N}}_{ij}}(z'))$ for class encouragements $z$ and $z'$.

\subsection{The Principal Stratification Approach}
\label{sec:PS}

%In the potential outcomes approach, causal effects are defined as comparison of potential outcomes on a common set of units, such as a comparison of the means of $Y_{ij}(z)$ and $Y_{ij}(z')$ for $i=1, \ldots, N_j$ and $j=1,\ldots,J$. 
In clustered encouragement designs with individual noncompliance, we are often interested in causal effects defined accounting for the compliance behavior. We deal with this issue using the principal stratification framework.

Principal stratification was proposed by \cite{Frangakis:Rubin:2002} as a general framework for the evaluation of treatment effects while addressing complications arising after the assignment to an encouragement or a treatment,  like non-compliance \citep[e.g.,][]{Hirano:2000,Frangakis:2002, Forastiere:2016} and truncation by death \citep[e.g.,][]{Mattei:2007}. 
%like non-compliance \citep[e.g.,][]{Frangakis:2002,	Frumento:2012}  and truncation by death \citep[e.g.,][]{Rubin:2006,Mattei:2007}. 
A principal stratification with respect to a post-assignment variable is a classification of units into latent groups, called ``principal strata'', defined by the joint potential values of that post-assignment variable under alternative levels of the assignment. Principal strata are latent groups because it is generally not possible to observe all the potential values of the post-assignment variable for any unit: it is only possible to observe the potential outcomes associated with the actual assignment, while the potential outcome under any alternative assignment is missing. Therefore, the specific stratum a unit belongs to is generally unknown.

In our study, we classify students into principal strata defined by their compliance behavior. Because the encouragement is a three-valued variable, $Z_j \in \left\lbrace 1,2,3\right\rbrace$, and the intermediate variable is a binary indicator for the visit Palazzo Vecchio by two months after randomization of the encouragement, $M_{ij} \in \left\lbrace 0,1\right\rbrace$, the principal stratification with respect to $M_{ij}$ classifies units into eight latent subgroups according to the joint potential values of $M_{ij}$ under each encouragement level: $M_{ij}(1)$, $M_{ij}(2)$ and $M_{ij}(3)$. More formally, let $G_{ij}=(M_{ij}(1),M_{ij}(2),M_{ij}(3))$, then  the principal strata are given by the possible values of $G_{ij}$:
%\textcolor{red}{$$G_{ij} := \{ij:  M_{ij}(1)=m_1, M_{ij}(2)=m_2, M_{ij}(3)=m_3\} \text{ with } m_1, m_2, m_3 \in \{0,1\}.$$}
$(m_1,m_2,m_3):= \{ij:  M_{ij}(1)=m_1, M_{ij}(2)=m_2, M_{ij}(3)=m_3\} \text{ with } m_1, m_2, m_3 \in \{0,1\}.$
 We relabel the possible values of $G_{ij}$ as $m_1m_2m_3$, so
%Let $G_{ij}$ denote the indicator for principal stratum membership for  student $i$ in class $j$: 
$G_{ij} \in \{ 111, 011, 101, 110, 001, 010, 100, 000 \}$. We name the eight principal strata as shown in Table \ref{tab:stratumlabel}.

\begin{table}[ht]
	\centering
	\caption{Principal strata and their labels}
	\label{tab:stratumlabel}
	\begin{tabular}{ccccl}
		\hline\noalign{\smallskip}
		$M_{ij}(1)$ & 	$M_{ij}(2)$& 	$M_{ij}(3)$& \textit{Principal Stratum  } & \textit{Stratum Label} \\ 
		\noalign{\smallskip}\hline\noalign{\smallskip}
		1 & 1 & 1 & $G = 111$ & \textit{Always Takers} \\
		0 & 1 & 1 & $G=011$& \textit{Presentation Compliers}  \\
		0 & 0 & 1 &$G=001$ & \textit{Reward Compliers} \\
		0 & 0 & 0 & $G=000$& \textit{Never Takers} \\ 
		1 & 0 & 0 &$G=100$ &\textit{Defiers}  \\ 		
		0 & 1 & 0 &$G=010$ &\textit{Defiers} \\ 		
		1 & 1 & 0 & $G=110$ &\textit{Defiers} \\ 		
		1 & 0 & 1 & $G=101$ &\textit{Defiers} \\ 
		\noalign{\smallskip}\hline
	\end{tabular}
\end{table}

\noindent As we can see in Table~\ref{tab:stratumlabel}, we refer to the union of principal strata $100,010,110,101$ as \textit{Defiers}. They are students who would visit Palazzo Vecchio under the flyer and/or presentation condition(s), but would not visit the museum under a stronger encouragement  (presentation and/or reward).
%For instance, students belonging to $110$ are a type of Defiers because they would visit Palazzo Vecchio if they received the presentation made by an art expert or even just the flyer, but would not visit the museum if they were also assigned to receive the extra-credit points towards their final school grade (as provided for in the reward encouragement).
In our application study, similar behaviors are implausible: if the flyer and/or the presentation is/are enough to motivate a student to visit Palazzo Vecchio, we see no reasons why receiving additional motivators should discourage the same visit. Therefore, we rule out the presence of Defiers making the following assumption:
\begin{assumption}(Monotonicity of Compliance).
	For all students $i$ in each class $j$, $M_{ij}(1) \le M_{ij}(2) \le M_{ij}(3)$.
	\label{ass:monotonicity}
\end{assumption}
\noindent Under monotonicity, $G_{ij} \in \{ 111, 011, 001, 000 \}$ for all $ij$, that is, each student $ij$ can belong to one of the following four groups:  \textit{Always Takers} (AT), \textit{Presentation Compliers} (PC), \textit{Reward Compliers} (RC) and \textit{Never Takers} (NT). \textit{Always Takers} and \textit{Never Takers} are students who, respectively, would perform, and would not perform the proposed visit to Palazzo Vecchio regardless of the encouragement. \textit{Presentation Compliers} are  students who would visit Palazzo Vecchio if they received the reward or at least attended the presentation. \textit{Reward Compliers} are students who would visit Palazzo Vecchio only upon receiving the reward. 

\subsection{Principal Causal Effects}
\label{sec:PCE}

The main advantage of the principal stratification approach is that it allows to define stratum-specific causal effects, named \textit{Principal Causal Effects} (PCEs), solving the complication of adjusting causal effects for the post-assignment variable, $M_{ij}$,  the actual visit to Palazzo Vecchio.
Because principal strata are not affected by encouragement assignment, PCEs are properly defined causal effects.

We focus on finite population average principal causal effects. Overall average principal causal effects of an encouragement condition, say $z$, versus another, say $z'$, are defined as follows:
\begin{equation}
	PCE_{g}(z \text{ vs } z')=\dfrac{ \sum_{ij: G_{ij}=g} Y_{ij}(z) - Y_{ij}(z')}{N_{g}}=\dfrac{ \sum_{ij: G_{ij}=g} Y_{ij}(z,S_{\mathcal{N}_{ij}}(z)) - Y_{ij}(z',S_{\mathcal{N}_{ij}}(z'))}{N_{g}}
	\label{eq:PCE}
\end{equation}
\noindent where $N_{g}$ is the total number of subjects in the sample belonging to the principal stratum $g \in \{000, 001, 011, 111\}$.

For Presentation and Reward Compliers, overall principal causal effects are \textit{associative}, in the sense that they measure effects on the outcome that are associative with effects on the intermediate variable, $M_{ij}$,  combining three effects: the \textit{pure encouragement effect}, the \textit{experience effect}, and the \textit{spillover effect} \cite[][]{Frangakis:Rubin:2002,Forastiere:2019}. 
%various effects \cite[][]{Frangakis:Rubin:2002}. Specifically, PCEs for these strata can be made of three components: the \textit{pure encouragement effect}, the \textit{experience effect}, and the \textit{spillover effect} \cite[][]{Forastiere:2019}. 
Pure encouragement effects measure changes in the individual primary outcome, i.e., number of museum visits in the follow-up period, directly induced by the encouragement.
Experience effects measure changes in the individual outcome that originate from individual treatment uptake, i.e., from visiting Palazzo Vecchio first-hand.
Finally, spillover effects measure changes in the individual outcome ascribable to mechanisms of interference sparked by friends performing the encouraged visit to Palazzo Vecchio in the first place. Spillover effects can be interpreted as channeled effects: they measure the part of the encouragement effects mediated through the proportion of friends visiting Palazzo Vecchio.

For Never Takers and Always Takers, overall principal causal effects are \textit{dissociative} \cite[][]{Frangakis:Rubin:2002}: because the visit to Palazzo Vecchio is -- by definition -- not affected by the encouragement for Never Takers and Always Takers, it follows that any change in the individual outcome may be ``only'' ascribed to  pure encouragement and  spillover effects \cite[][]{Forastiere:2019}.

In Table \ref{tab:PCE}, for each principal stratum, we resume the different causal pathways involved in the $PCEs$.
\begin{table}[ht]
	\centering
	\caption{Causal mechanisms underlying Principal Causal Effects}
	\label{tab:PCE}
	\begin{tabular}{l|l|l|l}
		\hline\noalign{\smallskip}
		\textit{Principal Stratum} &$PCE_g(3  \text{ vs } 2)$ & $PCE_g(2  \text{ vs } 1)$ & $PCE_g(3 \text{ vs } 1)$ \\
		\noalign{\smallskip}\hline\noalign{\smallskip}
		Never Takers &  Spill & Enc, Spill & Enc, Spill \\
		Always Takers & Spill & Enc, Spill & Enc, Spill\\
		Presentation Compliers &  Spill & Enc, Exp, Spill & Enc, Exp, Spill\\
		Reward Compliers & Exp, Spill & Enc, Spill & Enc, Exp, Spill\\
		\noalign{\smallskip}\hline
	\end{tabular}
	\begin{flushleft}
		\textit{Note}. ``Enc'', ``Exp'' and ``Spill'' stand for, respectively, pure encouragement effect, experience effect and spillover effect.
	\end{flushleft}
\end{table}
As we can see in Table \ref{tab:PCE}, we interpret the overall principal causal effects of reward ($z=3$) versus presentation ($z=2$) for Never Takers, Always Takers and Presentation Compliers as spillover effects, implicitly assuming that there do not exist pure encouragement and experience effects for these types of subjects.
Indeed, for Never Takers ($G_{ij}=000$), Always Takers ($G_{ij}=111$), and Presentation Compliers ($G_{ij}=011$), $M_{ij}(2)=M_{ij}(3)$, and thus, experience effects do not exist by definition. Moreover, for these types of students, we can reasonably assume that there exist no pure encouragement effect of reward ($z=3$) versus presentation ($z=2$). Under assignment to the reward, students get the reward only if they visit Palazzo Vecchio within two months after assignment; they do not get any reward if they visit art museums in the follow-up period but do not visit Palazzo Vecchio in the first place. Therefore, in terms of pure encouragement effects we can argue that the reward adds noting with respect to presentation.

\subsection{Disentangling Spillovers from PCEs Through Mediation}   
\label{sec: mediation} 
This section describes the core of our novel methodological proposal: using principal stratum mediation analysis to disentangle spillovers from the previous PCEs.
%\cite[see][for a review on mediation analysis]{Pearl:2014}. 

In principal stratum mediation analysis, focus is on disentangling the part of each principal causal effect that works through another variable, a mediator -- \textit{indirect principal causal effect}, from the causal effects of the assignment/treatment on the response net of the effect of the mediator -- \textit{direct principal causal effect}.
In our application study, we conduct principal stratum mediation analysis using the proportion of friends who visited Palazzo Vecchio as a mediator. Therefore, indirect principal causal effects measure changes in the response that can be ascribed to a variation in the summary of friends' compliance behavior by principal strata, and thus, they are interpretable as spillover effects. Formally, we introduce the concepts of principal natural direct and indirect effects and principal control direct effects:
\begin{eqnarray}
	\label{eq:NDE}
	NDE_g(z \text{ vs } z', S_{\mathcal{N}_{ij}}(z''))
	&=&\dfrac{\sum_{ij: G_{ij}=g}\left[Y_{ij}(z,S_{\mathcal{N}_{ij}}(z'')) - Y_{ij}(z',S_{\mathcal{N}_{ij}}(z''))\right]}{N_{g}}\\
	\nonumber\\
		\label{eq:NIE}
	NIE_g(z'', S_{\mathcal{N}_{ij}}(z) \text{ vs } S_{\mathcal{N}_{ij}}(z'))
	&=&\dfrac{\sum_{ij: G_{ij}=g}\left[Y_{ij}(z'',S_{\mathcal{N}_{ij}}(z)) - Y_{ij}(z'',S_{\mathcal{N}_{ij}}(z'))\right]}{N_{g}}\\
		\nonumber\\
		\label{eq:CDE}
		CDE_g(z \text{ vs } z',s^{*}_{\mathcal{N}_{ij}})&=&\dfrac{\sum_{ij: G_{ij}=g}\left[Y_{ij}(z,s^{*}_{\mathcal{N}_{ij}}) - Y_{ij}(z',s^{*}_{\mathcal{N}_{ij}})\right]}{N_{g}}
\end{eqnarray}	
for $z,z', z'' \in \{1,2,3\}$ with $z \neq z'$ and $z''$ set to $z$ or to $z'$, and $s^{*}_{\mathcal{N}_{ij}} \in [0,1]$ is a value of the mediator reasonably chosen by the researcher and not necessarily observed in the sample. In the application, we calculate these estimands for $z>z'$.

Principal natural direct effects measure average local effects (that is, for students of type $g$) of $z$ versus $z'$ on the response, the number of visits to art museums, not mediated through $S_{\mathcal{N}_{ij}}$, the proportion of friends visiting Palazzo Vecchio; in contrast, they consider the effect of $z$ versus $z'$ on the response intervening by fixing the proportion of friends visiting Palazzo Vecchio to the value it would have taken if the encouragement had been set to $z''$.
Principal natural direct effects embrace all causal channels other than spillovers, i.e., the pure encouragement effect and, where appropriate, also  experience effects. 

Principal natural indirect effects fix the encouragement to the value $z''$ and then
measure average local effects (that is, for students of type $g$) %of \textcolor{red}{$S_{\mathcal{N}_{ij}}(z)$ vs $S_{\mathcal{N}_{ij}}(z')$} 
on the response, the number of visits to art museums, mediated through the proportion of friends visiting Palazzo Vecchio, $S_{\mathcal{N}_{ij}}$; that is, intervening by setting $S_{\mathcal{N}_{ij}}$ to what it would have been if the encouragement were $z$, $S_{\mathcal{N}_{ij}}(z)$, in contrast to what it would have been if the encouragement were $z'$, $S_{\mathcal{N}_{ij}}(z')$. Principal natural indirect effects provide information on spillover effects.

Principal controlled direct effects measure local effects on the response not mediated through the proportion of friends visiting Palazzo Vecchio, as principal natural direct effects, but they compare potential outcomes for students belonging to the same principal stratum under two encouragements, $z$ and $z'$, setting the mediator, $S_{\mathcal{N}_{ij}}$, to some pre-fixed value $s^{*}_{\mathcal{N}_{ij}}$.

It is worth noting that ``Natural Direct Effects'' (NDE), ``Natural Indirect Effects'' (NIE), and ``Controlled Direct Effects'' (CDE), defined for the whole population, rather than for latent sub-populations such as the principal strata, are the causal estimands usually considered in mediation analysis \cite[e.g.,][]{RobinsGreenland:1992, Pearl:2001}.

Note that, as in the classical mediation setting overall average causal effects can be decomposed into the sum of natural direct and indirect effects, each average overall principal causal effect can be decomposed into the sum of a principal natural direct effect and a principal natural indirect effect: $PCE_g(z \text{ vs } z') = NDE_g(z \text{ vs } z', S_{\mathcal{N}_{ij}}(z)) \; + \; NIE_g(z',S_{\mathcal{N}_{ij}}(z') \text{ vs } S_{\mathcal{N}_{ij}}(z)) =  NDE_{g}(z' \text{ vs } z, S_{\mathcal{N}_{ij}}(z'))\; + \; NIE_{g}(z,S_{\mathcal{N}_{ij}}(z) \text{ vs } S_{\mathcal{N}_{ij}}(z'))$.
%\begin{eqnarray*}
%	PCE_g(z \text{ vs } z') &=& NDE_g(z \text{ vs } z', S_{\mathcal{N}_{ij}}(z)) \; + \; NIE_g(z',S_{\mathcal{N}_{ij}}(z') \text{ vs } S_{\mathcal{N}_{ij}}(z)) \\
%	&=&  NDE_{g}(z' \text{ vs } z, S_{\mathcal{N}_{ij}}(z'))\; + \; NIE_{g}(z,S_{\mathcal{N}_{ij}}(z) \text{ vs } S_{\mathcal{N}_{ij}}(z'))
%	%\label{eq:PCEdecomposition}
%\end{eqnarray*}

We also introduce new causal estimands, which we call principal   controlled spillover effects, defined as follows:
\begin{equation}
	\label{eq:CSE}
	CSE_g(z, s^{\ast\ast\ast}_{\mathcal{N}_{ij}} \text{ vs } s_{\mathcal{N}_{ij}}^{\ast \ast}) = \dfrac{\sum_{ij: G_{ij}=g}\left[Y_{ij}(z,s_{\mathcal{N}_{ij}}^{\ast \ast \ast}) - Y_{ij}(z,s_{\mathcal{N}_{ij}}^{\ast\ast})\right]}{N_{g}}
\end{equation}
where $z  \in \{1,2,3\}$  and   $s_{\mathcal{N}_{ij}}^{\ast\ast}$ and $s_{\mathcal{N}_{ij}}^{\ast\ast\ast} \in [0,1]$ are values of the proportion of friends visiting Palazzo Vecchio.
Principal controlled spillover effects measure local effects on the response for students belonging to the same principal stratum and exposed to the encouragement level $z$, due to a change in the proportion of friends visiting Palazzo Vecchio from a pre-fixed value $s_{\mathcal{N}_{ij}}^{\ast \ast \ast}$ to another pre-fixed value $s_{\mathcal{N}_{ij}}^{\ast \ast}$. Therefore, principal controlled spillover effects provide information on spillover effects.
	
As discussed in Section \ref{sec:PCE}, experience effects cannot exist for Always Takers and Never Takers. Therefore, for these two principal strata, principal natural and controlled direct effects provide information on pure encouragement effects, and principal natural indirect effects and principal controlled spillover effects provide information on spillover effects.

For Presentation Compliers and Reward Compliers, in general, there exist experience effects, and thus, overall principal causal effects for these principal strata combine encouragement, experience and spillover effects. Therefore, for Presentation Compliers and Reward Compliers, principal natural indirect effects and principal controlled spillover effects are interpretable as spillover effects, and principal natural and controlled direct effects entangle pure encouragement and experience effects. Table \ref{tab:NDENIE} summarizes the information on the causal pathways provided by principal natural and controlled direct effects. 
\begin{table}[ht]
	\centering
	\caption{Principal natural direct effects and principal controlled direct effects: an interpretation}
	\label{tab:NDENIE}
	\resizebox{\columnwidth}{!}{%
		\begin{tabular}{l|l|l|l}
			\hline\noalign{\smallskip}
			\textit{Principal Stratum } $(g)$ &
			\makecell{$NDE_g(3 \text{ vs } 2, S_{\mathcal{N}_{ij}}(z''))$\\ $CDE_g(3 \text{ vs } 2, s^{\ast}_{\mathcal{N}_{ij}})$} & 
			\makecell{$NDE_g(2 \text{ vs } 1, S_{\mathcal{N}_{ij}}(z''))$\\ $CDE_g(2 \text{ vs } 1, s^{\ast}_{\mathcal{N}_{ij}})$} & 
			\makecell{$NDE_g(3 \text{ vs } 1, S_{\mathcal{N}_{ij}}(z''))$\\ $CDE_g(3 \text{ vs } 1, s^{\ast}_{\mathcal{N}_{ij}})$} \\
			\noalign{\smallskip}\hline\noalign{\smallskip}
			Never Takers & 0 & Enc & Enc \\
			Always Takers & 0 & Enc & Enc \\
			Presentation Compliers & 0 & Enc, Exp & Enc, Exp\\
			Reward Compliers & Exp & Enc & Enc, Exp\\
			\noalign{\smallskip}\hline
		\end{tabular}
	}
	\begin{flushleft}
		\textit{Note}. ``Enc'' and ``Exp''  stand for   pure encouragement effect experience effect, respectively.
	\end{flushleft}
\end{table}
Note that there is no encouragement and experience effect for Never Takers, Always Takers and Presentation Compliers when encouragement levels $2$ and $3$ are compared, and thus, $NDE_g(3 \text{ vs } 2, S_{\mathcal{N}_{ij}}(z''))$ and $CDE_g(3 \text{ vs } 2, s^\ast_{\mathcal{N}_{ij}})$ for $g \in \{000,111,011\}$ are zero.

\subsection{Ignorability Assumptions}
Our study is a CED, and thus, the encouragement is randomly assigned by design, implying that the encouragement is independent of all the potential outcomes and covariates. Formally,
\begin{assumption} (Randomization of the encouragement) \label{ass:ced}\\
	%	$$Z_j \independent M_{ij}(1),  M_{ij}(2), M_{ij}(3), S_{\mathcal{N}_{ij}}(1),  S_{\mathcal{N}_{ij}}(2), S_{\mathcal{N}_{ij}}(3), \{Y_{ij}(1, s_{\mathcal{N}_{ij}})\}_s, \{Y_{ij}(2, s_{\mathcal{N}_{ij}})\}_s, \{Y_{ij}(3, s_{\mathcal{N}_{ij}})\}_s, \bm{X}_{ij}$$
		$Z_j \independent \{\{M_{ij}(z)\}_z, \{S_{\mathcal{N}_{ij}}(z)\}_z,  \{Y_{ij}(z, s_{\mathcal{N}_{ij}})\}_{z,s_{\mathcal{N}_{ij}}}, \bm{X}_{ij}\}$ for $z \in\{1,2,3\}$ and $s_{\mathcal{N}_{ij}} \in [0,1]$
\end{assumption}

Under monotonicity (Assumption \ref{ass:monotonicity}), there exist four principal strata and most observed groups defined by the encouragement actually received and the observed value of $M_{ij}$ host mixtures of  principal strata as shown in Table \ref{tab:obsgroups}.

\begin{table}[ht]
	\centering
	\caption{Observed groups of students and possible principal strata: NT=000, RC=001, PC=011, AT=111}
	\label{tab:obsgroups}
	\begin{tabular}{ccccccl}
		\hline\noalign{\smallskip}
		$Z_j$ & $M_{ij}$  & No. & $\%|Z_j=z$ & $Mean(Y_{ij})$ & $P(Y_{ij}>0)$ & \textit{Possible principal Strata} \\
		\noalign{\smallskip}\hline\noalign{\smallskip}
		1 & 0 & 86 & 96.6\% & 1.35 & 0.34 &  PC, RC, NT \\
		1 & 1 & 3 & 3.4\% & 5.67 & 1.00 & AT\\	
		2 & 0 & 77 & 86.5\% & 4.58 & 0.87 &  RC, NT\\
		2 & 1 & 10 & 11.2\% & 2.90 & 0.80 & AT, PC\\
		3 & 0 & 50 & 55.6\% & 3.18 & 0.88 & NT\\
		3 & 1 & 40 & 44.4\% & 2.78 & 0.95 & AT, PC, RC\\
		\noalign{\smallskip}\hline	
	\end{tabular}	
\end{table}

Randomization guarantees that principal stratum membership has the same distribution in all encouragement arms. Therefore, under randomization and monotonicity, we can  non-parametrically point identify principal stratum proportions. Let $\pi_{m_1m_2m_3}$ denote the proportion of units belonging to principal stratum $m_1m_2m_3=\{ij: M_{ij}(1)=m_1, M_{ij}(2)=m_2, M_{ij}(3)=m_3\}$, that is, $\pi_{m_1m_2m_3} := P(G_{ij} = m_1m_2m_3)$. We have
$\pi_{100} = \pi_{010} = \pi_{110} = \pi_{101} = 0$;
$\pi_{111}=P(M_{ij}=1 \mid Z_j=1); \quad \pi_{000} = P(M_{ij}=0\mid Z_j=3)$;
$\pi_{001} = P(M_{ij}=0\mid Z_j=2)-\pi_{000} = P(M_{ij}=1\mid Z_j=3)-P(M_{ij}=1\mid Z_j=2)$; and
$\pi_{011} = P(M_{ij}=1\mid Z_j=2)-\pi_{111} = P(M_{ij}=1\mid Z_j=3)-\pi_{111}-\pi_{001}$.
The method-of-moments estimates of the principal stratum proportions and also Table \ref{tab:obsgroups}) are as follows (standard errors in parenthesis): $\hat{\pi}_{111}=0.034\; (0.019)$; $\hat{\pi}_{000}=0.556\; (0.052)$; $\hat{\pi}_{001}=0.329\; (0.062)$; $\hat{\pi}_{011}=0.081\; (0.039)$; suggesting that there are more than 50\% Never Takers, and less than 3.5\% Always Takers. About 1/3 of students are expected to be Reward Compliers and 8.1\% of students are expected to be Presentation Compliers.

Although we can point identify principal stratum proportions, principal stratum membership is only partially observed, and thus, overall principal causal effect\textbf{s}, $PCE_g(z \text{ vs } z')$, are not non-parametrically point identified without additional assumptions. Inference on principal natural direct and indirect effects (Equations \ref{eq:NDE} and \ref{eq:NIE}) and principal controlled direct and spillover effects (Equations \ref{eq:CDE} and \ref{eq:CSE}) is even more challenging, because these estimands involve potential outcomes of the form $Y_{ij}(z, s_{\mathcal{N}_{ij}})$ and $Y_{ij}(z, S_{{\mathcal{N}}_{ij}}(z'))$, which are never observed in this specific experiment for some students. Some additional structural assumptions that allow us to extrapolate information on this type of potential outcomes from at least potentially observable data is required. Here we invoke  ``latent''  ignorability assumptions of the mediator, that is,  ignorability assumptions of the mediator conditional on principal stratum membership.

\begin{assumption} (Latent ignorability of the mediator (I)). For each unit $i$ in cluster $j$, 	for all $z \in \{1,2,3\}$ and $s_{\mathcal{N}_{ij}} \in [0,1]$,
		$Y_{ij}(z, s_{\mathcal{N}_{ij}}) \independent S_{\mathcal{N}_{ij}} \mid Z_j, G_{ij},\bm{X}_{ij}$;
	\label{ass:ign_med1}
\end{assumption}

\begin{assumption} (Latent ignorability of the mediator (II)). For each unit $i$ in cluster $j$, 	for all $z, z' \in \{1,2,3\}$ and $s_{\mathcal{N}_{ij}} \in [0,1]$, 
	$Y_{ij}(z, s_{\mathcal{N}_{ij}})  \independent S_{\mathcal{N}_{ij}}(z') \mid G_{ij},\bm{X}_{ij}$.

	\label{ass:ign_med2}
\end{assumption}
In our application study, Assumption \ref{ass:ign_med1} means that potential outcomes of the form $Y_{ij}(z, s_{\mathcal{N}_{ij}})$  for future visits to other  museums of a student $i$ in class $j$ are independent of the decision to visit Palazzo Vecchio of subject $i$'s friends in class $j$ conditional on  student $i$'s encouragement status, covariates and compliance status. 
Assumption \ref{ass:ign_med2} implies that $Y_{ij}(z, s_{\mathcal{N}_{ij}})$ and
the decision to visit Palazzo Vecchio of subject $i$'s friends in class $j$ under encouragement $z'$, $S_{\mathcal{N}_{ij}}(z')$, are independent conditional on student $i$'s covariates and compliance status.

If the compliance status were fully observed for each unit, Assumption \ref{ass:ign_med1} and Assumption \ref{ass:ign_med2} would be the standard ignorability assumptions on the mediator usually invoked in the causal mediation literature \cite[see, e.g.,][for a review]{VanderWeeleVansteelandt:2009, TenHave:2012}. 
Therefore, we could identify principal controlled direct and spillover effects under  Assumptions \ref{ass:ced} and \ref{ass:ign_med1} and principal natural direct and indirect effects 	under  Assumptions \ref{ass:ced}, \ref{ass:ign_med1} and  \ref{ass:ign_med2} (see the online Appendix for details).
%
%Therefore, under  Assumptions \ref{ass:ced} and \ref{ass:ign_med1}, we could identify principal controlled direct and spillover effects using the following relationship between the conditional mean of the potential outcomes, $Y_{ij}(z, s_{\mathcal{N}_{ij}})$, given principal stratum membership and covariates and the conditional mean of the observed outcomes, $Y_{ij}$, given principal stratum membership, encouragement status, mediator and covariates:
%\begin{equation}\label{eq:ce}
%	E\left[Y_{ij}(z, s_{\mathcal{N}_{ij}}) \mid G_{ij}=g, \textbf{X}_{ij}\right] = E\left[Y_{ij} \mid Z_j=z, S_{\mathcal{N}_{ij}} =s_{\mathcal{N}_{ij}},G_{ij}=g,  \textbf{X}_{ij}\right]
%\end{equation}	
%Under  Assumptions \ref{ass:ced}, \ref{ass:ign_med1} and  \ref{ass:ign_med2},	
%we could identify principal natural direct and indirect effects using the mediation formula \cite[][]{Pearl:2001}:
%\begin{eqnarray}\label{eq:mf}
%	\lefteqn{
%		E\left[Y_{ij}(z, S_{\mathcal{N}_{ij}}(z')) \mid G_{ij}=g, \textbf{X}_{ij}\right] =}\\&&
%	\int E\left[Y_{ij} \mid Z_j=z, S_{\mathcal{N}_{ij}} =s_{\mathcal{N}_{ij}},G_{ij}=g,  \textbf{X}_{ij}\right] f_{S_{ij}}(s_{\mathcal{N}_{ij}} \mid  Z_j=z',G_{ij}=g, \textbf{X}_{ij}) \nonumber
%\end{eqnarray}
%Because the compliance status is only partially observed, we cannot directly use these relationships.
Because the compliance status is only partially observed,   Assumptions \ref{ass:ced}, \ref{ass:ign_med1} and  \ref{ass:ign_med2} do not solve identification issues.
We deal with inferential issues arising from the latent nature of principal stratum membership using a model-based Bayesian approach, which does not need full identification \cite[][]{Imbens:Rubin:1997}.
A Bayesian model-based principal stratification analysis requires the specification of two sets of models: one for the principal strata membership given the covariates, and one for the outcomes conditional on the covariates and the principal strata. In addition, a prior distribution for the model parameters must be specified.

\section{Analyses}

\subsection{Parametric Models}
In our setting, randomization is at cluster-level, and thus, a within-class correlation among individuals may arise from reciprocal influence or from other unobserved common factors. In particular, students belonging to the same class are likely to show resemblance in terms of both  compliance behavior and final outcomes.
We account for this systematic unexplained variation within classes introducing a hierarchical structure, by specifying varying-intercept models for the compliance status and for the outcome. 

\subsubsection*{Principal Strata Model}
We specify a  multinomial logit random effect model for principal stratum membership conditional on covariates:
\begin{equation}
    \begin{gathered}
    \log\left(\dfrac{P(G_{ij}=g | \bm{X_{ij}})}{P(G_{ij}=111  | \bm{X_{ij}})}\right) = \gamma_g+\bm{\delta_g^{T}}\bm{X_{ij}} + a_{j}
   \qquad \hbox{with }
    a_{j} \sim \mathit{N}\left(0 , \sigma_a\right),\\
    \end{gathered}
    \end{equation}
where $g \in \{000,001,011\}$ (the stratum $111$ is chosen as the reference category) and $a_{j}$ are the random intercepts. %Other models can be an alternative to the multinomial logit random effect model: in its place, for example, we could have assumed a hierarchical ordered probit model. However, since the model is specified as a flexible function of the covariates, the results are not expected to change based on the model chosen. 

\noindent The vector $\bm{X}_{ij}$ includes the individual background characteristics introduced in Table \ref{tab:covariates}: \textit{Male} (1/0), \textit{Already
	visited Palazzo Vecchio} (1/0), \textit{No. of museums visited previous year}, \textit{GPA} (continuous on a 1-10
points scale), \textit{Interest in human sciences} (1/0) and \textit{Parental education} (1 if at least one of the parents completed university, 0 otherwise). In addition, it includes a network variable indicating whether friends already visited Palazzo Vecchio prior to the experiment or not.

\subsubsection*{Potential Outcome Models}
We specify Zero-Inflated Poisson (ZIP) models for the potential outcomes conditional on principal stratum membership and covariates. A ZIP model assumes that subjects are a mixture of two types.  For subjects of first type the outcome is always zero. 
For subjects of the second type, the outcome follows an usual Poisson distribution, which can produce the zero outcome or some other. 
Let $\phi_{ij,g,z}$ denote the conditional probability given the covariates that an individual with compliance behaviour $g$ under assignment $z$ is of the first type,   and let $\mu_{ij,g,z}$ denote the Poisson mean, which we model   using a hierarchical generalized linear model with a log link:
%Let $\phi_{ij,g,z}$ denote the probability that an individual with compliance behaviour $g$ under assignment $z$ is of the first type, $\phi_{ij, g,z} = Pr(\hbox{subject $i$ in class $j$ is of the first type under assignment $z$} \mid G_{ij}=g, \bm{X}_{ij})$, and let $\mu_{ij,g,z}$ denote the Poisson mean, which we model   using a hierarchical generalized linear model with a log link:
\begin{equation}\label{eq:ZIP}
Pr\left(Y_{ij}\big(z,S_{\mathcal{N}_{ij}}(z)\big) =y \mid G_{ij}=g,\bm{X}_{ij}\right)=\\
\begin{cases}
\phi_{ij,g,z} + (1-\phi_{ij,g,z}) \cdot Pois\Big(0; \mu_{ij, g,z} \Big) &\text{if } y=0\\
(1-\phi_{ij,g,z}) \cdot Pois\Big(y; \mu_{ij, g,z} \Big) & \text{if } y>0\\
\end{cases}
\end{equation}
where
$$
Pois\Big(y; \mu_{ij, g,z} \Big) = \dfrac{e^{-\mu_{ij,g,z}}\mu_{ij,g,z}^{y}}{y!}\qquad y =0,1,2, \ldots
$$

\noindent In the ZIP models for the potential outcomes, we condition on the same individual and friends' background characteristics we include in the principal strata model and on an additional network variable indicating the number of friends' past museums visits. We include the latter because we can reasonably believe that having museum goers friends may influence the individual museums attendance.

As we discussed in Section~\ref{sec:PCE}, we can reasonably assume that there is no encouragement effects and experience effects for Never Takers, Always Takers and Presentation Compliers in the comparison of potential outcomes under reward ($z=3$) versus presentation ($z=2$) (see Table \ref{tab:NDENIE}). This assumption implies that for $g \in \{000,111,011\}$, 
$\phi_{ij,g,2}=\phi_{ij,g,3}$ and $\mu_{ij,g,2}=\mu_{ij,g,3}$, which, in turn, implies that $\alpha_{g,2}=\alpha_{g,3}$,  $\beta_{g,2}^{(S)}=\beta_{g,3}^{(S)}$ and $\bm{\beta}_{g,2}^{(X)}=\bm{\beta}_{g,3}^{(X)}$.

For the sake of parsimony, we assume that the probabilities of extra zero\textbf{s} with respect to the Poisson distribution, $\phi_{ij, g,z}$, depend on the encouragement
and the principal stratum membership, but not on covariates: $\phi_{ij, g,z} =\phi_{g,z}$. Moreover,  we impose the following prior equalities of the slope coefficients  and the standard deviations of the random intercepts in the Poisson regressions:
$\beta_{g,z}^{(S)} \equiv  \beta_{z}^{(S)}$,
 %so that the regression coefficients for $S_{\mathcal{N}_{ij}}(z)$  are only encouragement-specific; 
for $g \in \{000,001,011,111\}$ and $z \in \{1,2\}$, 
 $\bm{\beta}_{g,z}^{(X)} \equiv \bm{\beta}^{(X)}$ for all $z \in \{1,2,3\}$ and $g \in \{000,001,011,111\}$, and $\sigma_{b,z} \equiv \sigma_{b}$, for all $z \in \{1,2,3\}$.

\subsection{Bayesian Inference}
We use a Bayesian approach for inference, under the assumption that  parameters are a priori independent. We specify  relative weakly informative prior distributions, using normal priors
for the coefficients of the principal strata model and of the potential outcome models and half-normal prior distributions for the standard deviations $\sigma_{a}$ and $\sigma_{b}$ of the random intercepts $a_{j}$ and $b_{j}$  entering the principal strata model and the potential outcome models \cite[][]{Gelman:2006}.
The posterior distribution is approximated via Stan, a Bayesian programming language that implements a variant of Hamiltonian Monte Carlo \citep{Stan:2017}, and by using a data augmentation approach. The latter involves, at each iteration, the imputation of the missing potential outcomes for $Y_{ij}$ and $M_{ij}$ for each unit. Then, the effects are estimated as function of the observed and missing information. See the online appendix for details on prior specifications, likelihood function and computational details.

\section{Results}

\subsection{Principal Strata}
%We first focus on results on principal strata size and characteristics. 
Table \ref{tab:membership} shows some summary statistics for the posterior distribution of the principal strata membership. The estimated proportions of the principal strata suggest there are more than 58\% Never Takers, more than 30\% Reward Compliers, less than 10\% Always Takers and only 2\% Presentation Compliers.
Therefore, the majority of students do not visit Palazzo Vecchio irrespective of the encouragement, and a large minority of students is induced to visit Palazzo Vecchio if a reward is promised. Next to these two large strata, there are other more numerically negligible minorities for whom any encouragement works or for whom the presentation constitutes a sufficient motivator.
\begin{table}[ht]
	\centering
	\caption{Summary statistics of the posterior probabilities of principal stratum membership}
	\label{tab:membership}
	\begin{tabular}{lcccccc}
		\hline\noalign{\smallskip}
		& \textit{Mean} & \textit{SD} & \textit{2.5\%} & \textit{5\%} & \textit{95\%} & \textit{97.5\%} \\ 
		\noalign{\smallskip}\hline\noalign{\smallskip}
		Always Takers & 0.095 & 0.025 & 0.053 & 0.056 & 0.135 & 0.143 \\ 
		Presentation Compliers & 0.020 & 0.022 & 0.000 & 0.000 & 0.068 & 0.079 \\ 
		Reward Compliers & 0.301 & 0.049 & 0.203 & 0.218 & 0.376 & 0.395 \\ 
		Never Takers & 0.585 & 0.037 & 0.515 & 0.526 & 0.647 & 0.658 \\ 
		\hline
	\end{tabular}
\end{table}

It is worth nothing that the posterior means for the principal stratum proportions are slightly different from the  method-of-moment estimates, especially for the proportion of Always Takers and Presentation Compliers. These difference are mainly attributable to the fact that the method-of-moment estimates do not account for  individual and network level student's pre-experimental characteristics, which may affect the estimates given the observed imbalance across encouragement groups. Nevertheless, results are consistent, in the sense that both the unadjusted method-of-moment estimates and the adjusted model-based Bayesian estimates show evidence that the vast majority of students are either Never Takers or Reward Compliers and a small proportion of students are either Always Takers or Presentation Compliers.

An appealing advantage of the model-based Bayesian approach we use is that it easily allows us to investigate the distribution of baseline characteristics within each principal stratum, which may provide precious information on principal stratum members and on the propensity of members' friends to belong to the same latent principal stratum.

\begin{figure}[H]
	\centering
	\caption{Distribution of covariates by principal strata}
	\includegraphics[width=\columnwidth]{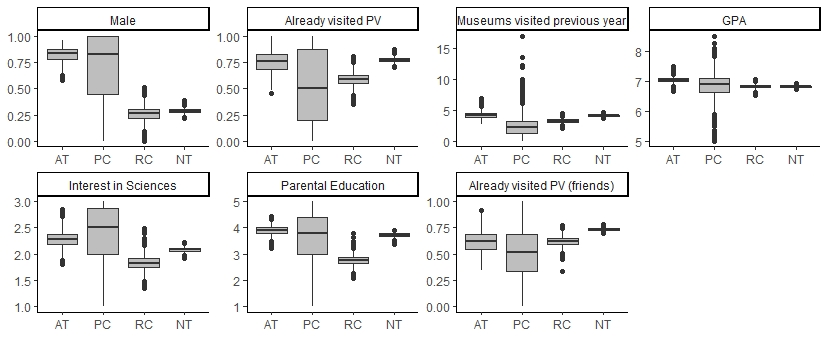}
	\label{fig:boxplots}
\end{figure}

Figure \ref{fig:boxplots} shows the boxplots of the posterior distributions of individual and friends' covariates in each principal stratum. All strata comprise students with similar GPAs, but they are somewhat different in terms of the other characteristics. 

The small clutch of Always Takers is mostly composed of male students from fairly educated families, whose scholastic interests are slightly more in the sciences than in the humanities, and that visited some museums during their spare time in the pre-experimental period. They are very likely to have already visited Palazzo Vecchio. Also, their friends are quite likely to have visited Palazzo Vecchio previously. 

The stratum of Never Takers is populated by a large majority of females, mostly from educated families, equally interested in the sciences and humanities. They have previously visited some museums in their free time. Both Never Takers and their friends are likely to have already visited Palazzo Vecchio. 

The tiny stratum of Presentation Compliers mostly consists of males, from fairly educated families, with a marked preference for the sciences over the humanities. They are slightly less accustomed than others to spending their free time in museums. They are also less likely than the members of other strata, and also than their own friends, to have already visited Palazzo Vecchio.

Finally, the stratum of Reward Compliers mostly hosts female students from not very educated families, with a marked preference for the humanities over the sciences. They have already dedicated their free time to visiting some museum\textbf{s}, but they are slightly less likely than Never Takers and Always Takers to have already visited Palazzo Vecchio. Their friends are also slightly less likely to have visited Palazzo Vecchio previously than   Never Takers' and Always Takers' friends.

\begin{figure}[ht]
	\centering
	\caption{Boxplots of the posterior distributions of proportion of friends in each stratum for a student belonging to a given stratum}
	\includegraphics[width=\textwidth]{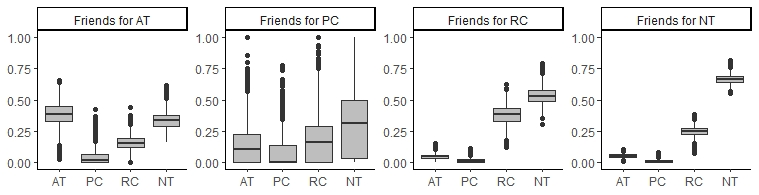}
	\label{fig:homophily}
\end{figure}

Figure \ref{fig:homophily} shows the boxplots of the posterior distributions of the proportions of friends in each stratum for a student belonging to a given stratum, providing   insights on 
the extent to which befriended students belong to the same principal stratum.
%We also look at the posterior distributions of the proportions of friends in each stratum for a student belonging to a given stratum. Figure \ref{fig:homophily} shows the boxplots of these distributions. These results allow us to examine to which extent befriended students belong to the same principal stratum. 
In a sense, this kind of analysis is informative about the existence of homophily in terms of the latent characteristics of the befriended students rather than, as it is usually done in the literature, in terms of their observable characteristics. As shown in Figure \ref{fig:homophily}, Always Takers are inclined to be befriended with other Always Takers and Never Takers, not so much with Reward Compliers and Presentation Compliers.
Presentation Compliers are equally likely to be befriended with the members of all the other strata, but it is more difficult for them to be friends with their stratum peers. Reward Compliers tend to have relationships with Never Takers and, to a lesser extent, with their peers. Finally, Never Takers are most likely to have ties with their peers and, to a far lesser extent, with Reward Compliers.

\subsection{PCEs and Principal Natural Direct and Indirect Effects}
Of the four strata described above, we now focus on the two largest ones, i.e., Never Takers and  Reward Compliers. For these two principal strata, Table \ref{tab:effects} reports summary statistics of the posterior distributions of the finite-population overall principal causal effects and principal natural direct and indirect effects on  count of museum visits during the follow-up period.
\begin{table}[ht]
	\centering
	\caption{Summary statistics of the posterior distributions of overall principal causal effects and principal natural direct and indirect effects for Never Takers and Reward Compliers}
	\label{tab:effects}
	\resizebox{\columnwidth}{!}{%
		\begin{tabular}{l|rrrrr|rrrrr}
			\noalign{\smallskip}\hline\noalign{\smallskip}
			&\multicolumn{5}{c}{Never Takers $(G_{ij}=000)$} & \multicolumn{5}{c}{Reward Compliers $(G_{ij}=001)$}\\
			Causal Estimand		& Mean & SD & 5\% & 95\% & Pr($>$0) & Mean & SD & 5\% & 95\% & Pr($>$0) \\ 
			\noalign{\smallskip}\hline\noalign{\smallskip}
			$PCE_g(2\text{ vs }1)$ & 1.65 & 1.04 & 0.07 & 3.35 & 0.96 & 4.73 & 1.42 & 2.66 & 7.14 & 1.00 \\ 
			$NDE_g(2\text{ vs }1, S_{\mathcal{N}_{ij}}(2))$ & 1.66 & 1.06 & -0.01 & 3.33 & 0.95 & 4.70 & 1.44 & 2.61 & 7.12 & 1.00 \\ 
			$NDE_g(2\text{ vs }1, S_{\mathcal{N}_{ij}}(1))$ & 1.65 & 1.05 & 0.03 & 3.39 & 0.95 & 4.53 & 1.53 & 2.32 & 7.12 & 1.00 \\ 
			$NIE_g(2, S_{\mathcal{N}_{ij}}(2)\text{ vs }S_{\mathcal{N}_{ij}}(1))$ & 0.00 & 0.21 & -0.36 & 0.35 & 0.51 & 0.20 & 0.48 & -0.59 & 0.96 & 0.66 \\ 
			$NIE_g(1, S_{\mathcal{N}_{ij}}(2)\text{ vs }S_{\mathcal{N}_{ij}}(1))$ & -0.01 & 0.20 & -0.32 & 0.33 & 0.46 & 0.02 & 0.25 & -0.28 & 0.46 & 0.45 \\
			\noalign{\smallskip}\hline\noalign{\smallskip}
			$PCE_g(3\text{ vs }2)$ & -0.03 & 0.28 & -0.50 & 0.43 & 0.45 & -2.39 & 1.29 & -4.56 & -0.46 & 0.02 \\ 
			$NDE_g(3\text{ vs }2, S_{\mathcal{N}_{ij}}(3))$ & 0.00 & 0.00 & 0.00 & 0.00 & 0.00 & -2.03 & 1.35 & -4.35 & 0.02 & 0.05 \\ 
			$NDE_g(3\text{ vs }2, S_{\mathcal{N}_{ij}}(2))$ & 0.00 & 0.00 & 0.00 & 0.00 & 0.00 & -2.32 & 1.49 & -4.74 & -0.04 & 0.05 \\ 
			$NIE_g(3, S_{\mathcal{N}_{ij}}(3)\text{ vs }S_{\mathcal{N}_{ij}}(2))$ & -0.03 & 0.28 & -0.50 & 0.43 & 0.45 & -0.07 & 0.60 & -1.13 & 0.78 & 0.50 \\ 
			$NIE_g(2, S_{\mathcal{N}_{ij}}(3)\text{ vs }S_{\mathcal{N}_{ij}}(2))$ & -0.03 & 0.28 & -0.50 & 0.43 & 0.45 & -0.36 & 0.60 & -1.32 & 0.67 & 0.27 \\ 
			\noalign{\smallskip}\hline\noalign{\smallskip}
			$PCE_g(3\text{ vs }1)$ & 1.62 & 1.03 & 0.03 & 3.31 & 0.95 & 2.34 & 1.04 & 0.69 & 3.99 & 0.99 \\ 
			$NDE_g(3\text{ vs }1, S_{\mathcal{N}_{ij}}(3))$ & 1.64 & 1.11 & -0.17 & 3.35 & 0.94 & 2.30 & 1.13 & 0.41 & 3.98 & 0.97 \\ 
			$NDE_g(3\text{ vs }1, S_{\mathcal{N}_{ij}}(1))$ & 1.65 & 1.05 & 0.03 & 3.39 & 0.95 & 2.40 & 1.33 & 0.43 & 4.58 & 0.98 \\ 
			$NIE_g(3, S_{\mathcal{N}_{ij}}(3)\text{ vs }S_{\mathcal{N}_{ij}}(1))$ & -0.03 & 0.30 & -0.53 & 0.45 & 0.46 & -0.07 & 0.62 & -1.14 & 0.81 & 0.50 \\ 
			$NIE_g(1, S_{\mathcal{N}_{ij}}(3)\text{ vs }S_{\mathcal{N}_{ij}}(1))$ & -0.02 & 0.49 & -0.63 & 0.81 & 0.40 & 0.03 & 0.41 & -0.43 & 0.73 & 0.41 \\ 
			\noalign{\smallskip}\hline
		\end{tabular}%
	}
\end{table}

Let us first focus on the contrast between Presentation and Flyer. 
We find evidence that principal causal effects of Presentation versus Flyer for Never Takers and Reward Compliers are positive. 
The posterior mean of the PCE for Never Takers is 1.65 (SD$ = 1.04$) and the 90\% credible interval only covers positive values; the posterior probability that this effects is positive is very high (96\%). The posterior mean of the PCE for Reward Compliers suggests an increase of 4.73 museum visits due to presentation versus flyer (SD$=$ 1.42 and 90\%CI = (2.66, 7.14)) and this principal causal effect takes on positive values with probability one.
Both results are consistent with those in \citet{Forastiere:2019}.  
Based on these results, Presentation appears as an effective encouragement towards attending museums in the future for Reward Compliers and, to a smaller extent, for Never Takers. 

As argued in Section \ref{sec:PCE}, overall PCEs may originate from both a direct and an indirect/spillover causal pathway.   Disentangling these pathways is our main contribution, moving a step forward with respect to the analyses in \citet{Forastiere:2019} and enriching the literature on causal inference with interference. When comparing the Presentation and the Flyer, the direct pathway, whose strength is evaluated by principal NDEs, channels only encouragement effects for both Never Takers and Reward Compliers. Instead, the indirect pathway, whose strength is evaluated by principal NIEs, may only channel spillovers. Principal natural direct effects of Presentation versus Flyer can be evaluated setting the proportion of friends visiting  Palazzo Vecchio to the value that it would have taken  either under the Flyer encouragement, $S_{\mathcal{N}_{ij}}(1)$, or under the Presentation encouragement, $S_{\mathcal{N}_{ij}}(2)$.
Similarly, principal natural indirect effects measure spillover effects on  future museum attendance arising from a change in the proportion of friends visiting  Palazzo Vecchio from $S_{\mathcal{N}_{ij}}(1)$, the value that $S_{\mathcal{N}_{ij}}$ would have taken under the Flyer encouragement ($z=1$)\textbf{,} to $S_{\mathcal{N}_{ij}}(2)$, the value that $S_{\mathcal{N}_{ij}}$ would have taken under the Presentation encouragement ($z=2$), by fixing the individual encouragement to either the Flyer level ($z=1$) or the Presentation level ($z=2$).
It is worthwhile to highlight that  $S_{\mathcal{N}_{ij}}(1)$ is equal to the proportion of Always Taker friends, because only Always Takers visit Palazzo Vecchio under the Flyer encouragement; analogously, $S_{\mathcal{N}_{ij}}(2)$ is equal to the proportion of friends that are either Always Takers or Presentation Compliers, because only Always Takers and Presentation Compliers visit Palazzo Vecchio under the Presentation encouragement.
Therefore, principal natural indirect effects quantify  the spillover on future museum attendance that would come from friends ready to visit Palazzo Vecchio under the Presentation but not under the Flyer, that is, from being exposed, in addition to Always Takers, to friends that react as Presentation Compliers.
Since the proportions of Always Takers and Presentation Compliers are small, $S_{\mathcal{N}_{ij}}(1)$ and $S_{\mathcal{N}_{ij}}(2)$  describe a friendship environment where a student is hardly exposed to friends that might send spillovers.

The posterior means of $NDE_g(2 \text{ vs } 1, S_{\mathcal{N}_{ij}}(2))$ and $NDE_g(2 \text{ vs } 1, S_{\mathcal{N}_{ij}}(1))$  are 1.66 and 1.65 for Never Takers $(g=000)$ and $4.70$ and $4.53$ for Reward Compliers $(g=001)$, and the corresponding posterior probabilities that these effects are positive are higher than 0.95. We interpret these results as evidence that for Never Takers and Reward Compliers there exist pure encouragement effects of Presentation versus Flyer.
The posterior means of $NIE_g(1,  S_{\mathcal{N}_{ij}}(2) \text{ vs }  S_{\mathcal{N}_{ij}}(1))$ and $NIE_g(2,  S_{\mathcal{N}_{ij}}(2) \text{ vs }  S_{\mathcal{N}_{ij}}(1))$ are very small for both Never Takers and Reward Compliers, and the corresponding 90\% posterior credible intervals cover zero, therefore there is no evidence that there exist spillover effects of Flyer versus Presentation for these type of subjects. Indeed, as we can see in Table \ref{tab:effects}, the size of the estimated principal natural direct effects for Never Takers and Reward Compliers is similar to the size of the overall principal causal effects. Therefore, causal effects for Never Takers and Reward Compliers are almost entirely due to the encouragement.

Let us now focus on the contrast between Reward ($z=3$) and  Presentation ($z=2$). For Never Takers, principal natural direct effects of Presentation versus Reward are zero by assumption (see the discussion in Sections ~\ref{sec:PCE} and \ref{sec: mediation}), therefore principal causal effects of $z=3$ versus $z=2$ for this type of subjects are interpretable as spillover effects. Consistently with the insights of \citet{Forastiere:2019}, we find no evidence that there exist statistically significant spillover effects of Presentation versus Reward for Never Takers: the posterior means of $NIE_{000}(2,  S_{\mathcal{N}_{ij}}(3) \text{ vs }  S_{\mathcal{N}_{ij}}(2))$ and $NIE_{000}(3,  S_{\mathcal{N}_{ij}}(3) \text{ vs }  S_{\mathcal{N}_{ij}}(2))$ are very close to zero and the corresponding 90\% posterior credible interval are spread around zero. For Reward Compliers, the posterior probabilities that overall principal causal effects of Presentation versus Reward are negative are close to 1, in line with the results previously presented in \citet{Forastiere:2019}. We interpret these overall principal causal effects for Reward Compliers as a mixture of experience and  spillover effects, which we propose to disentangle using principal natural direct and indirect effects. Specifically, we attribute principal natural direct effects of Presentation versus Reward for Reward Compliers solely to the experience (the visit at Palazzo Vecchio), because they visit Palazzo Vecchio only under Reward, and principal natural indirect effects to spillovers.
Summary statistics of the posterior distributions of these effects shown in Table \ref{tab:effects} suggest that experience effects account for almost all of the negative overall principal causal effect. Thus, Reward Compliers do not seem to have enjoyed their experience at Palazzo Vecchio very much, which may have even discouraged further museum visits in the follow-up period. If the effect of the Presentation itself is to enhance Reward Compliers' museum attendance, such effect is partially offset by the experience at Palazzo Vecchio. Spillovers are, instead, negligible. This conclusion on spillovers is particularly relevant because, contrary to the Presentation/Flyer contrast, the Reward is able to increase the proportion of friends visiting Palazzo Vecchio, by attracting Reward Compliers, who are about 1/3 of students and may act as senders of spillovers.

Finally, let us focus on the contrast between  Reward and Flyer. Overall principal causal effects of Reward versus Flyer amounts to the sum of the overall principal casual effects of Presentation versus Flyer and of Reward versus Presentation. We find evidence that  for both Never Takers and Reward Compliers, these principal causal effects are positive and mainly due to direct effects: the posterior distributions of principal natural direct effects closely overlap the posterior distributions of overall principal causal effects, and posterior distributions of overall principal natural indirect effects are concentrated on values close to zero and evenly spread around zero.

\subsection{Principal Controlled and Spillover Effects}
\begin{figure}[!ht]
	\centering
	\caption{Estimated CDEs}
	\includegraphics[width=0.244\textwidth]{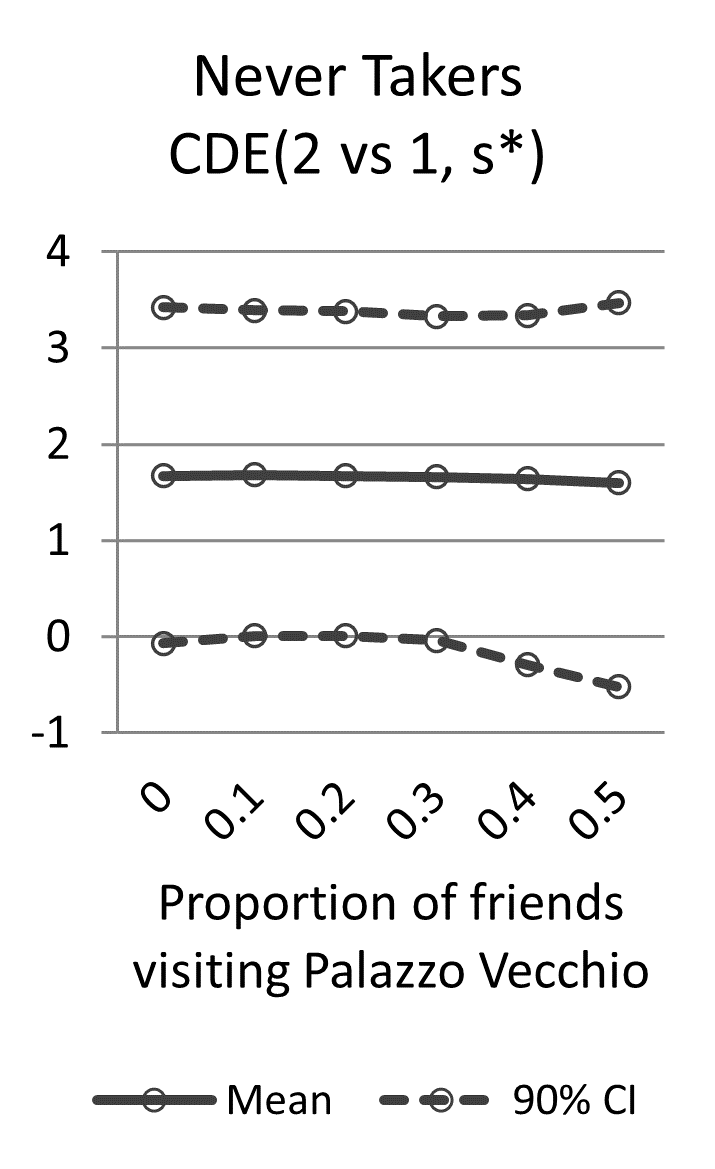}
	\includegraphics[width=0.244\textwidth]{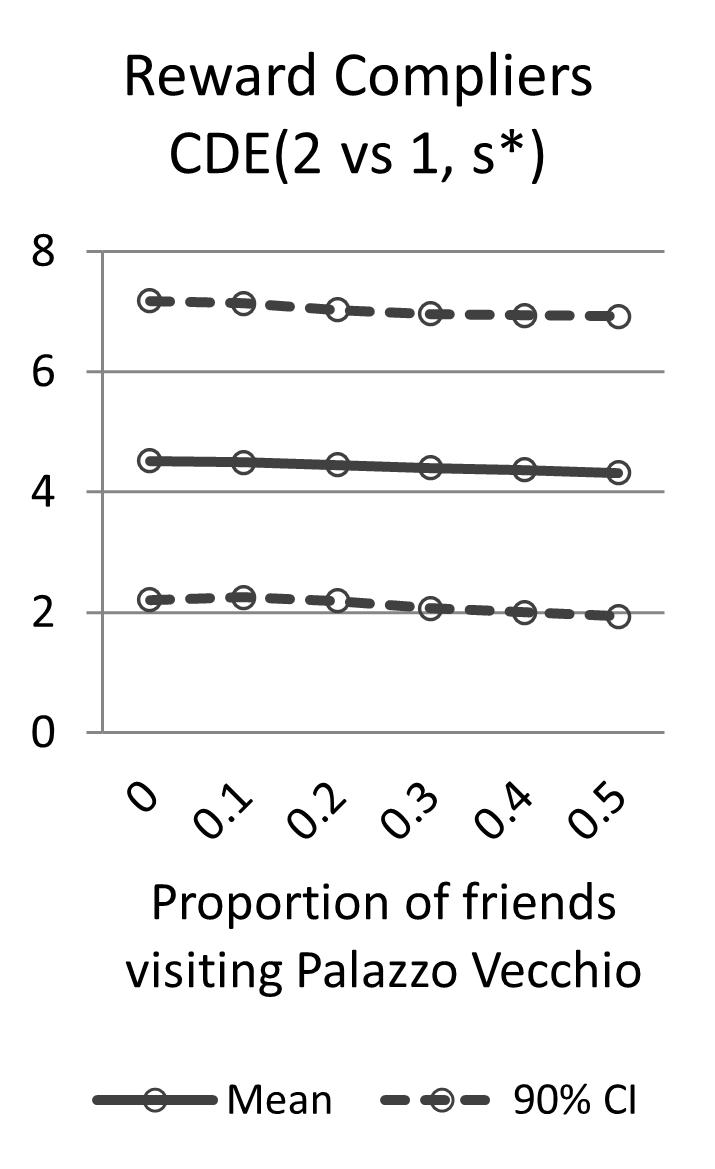}
	\includegraphics[width=0.244\textwidth]{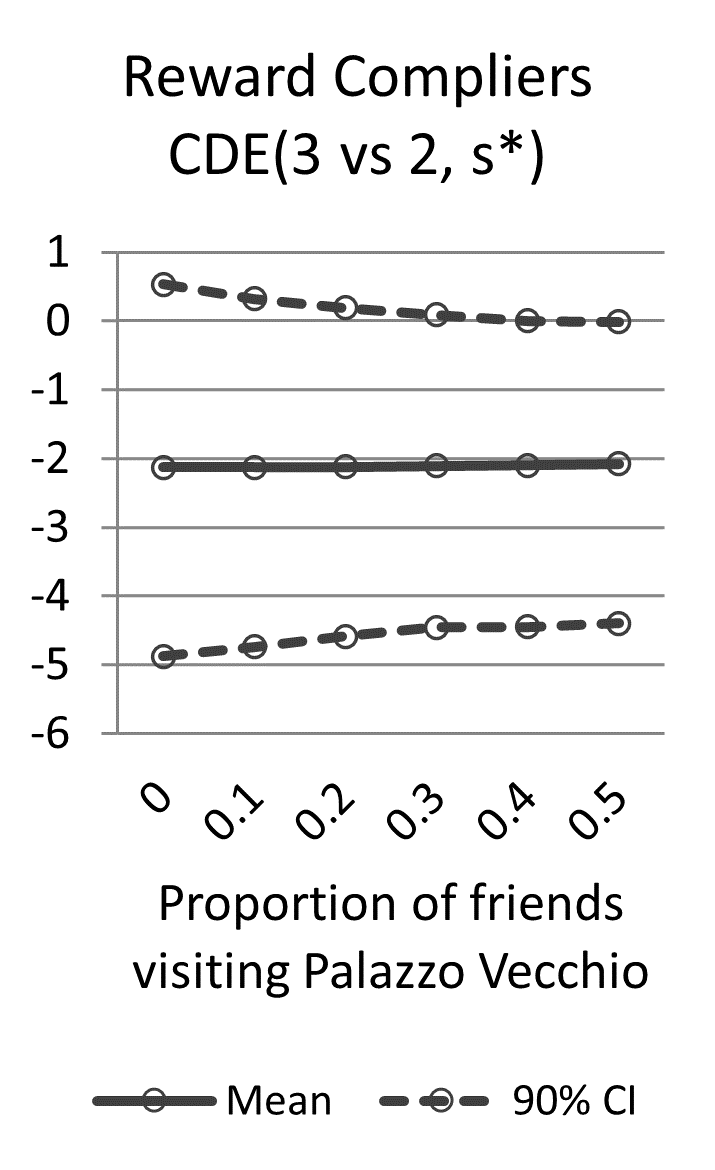}
	\includegraphics[width=0.244\textwidth]{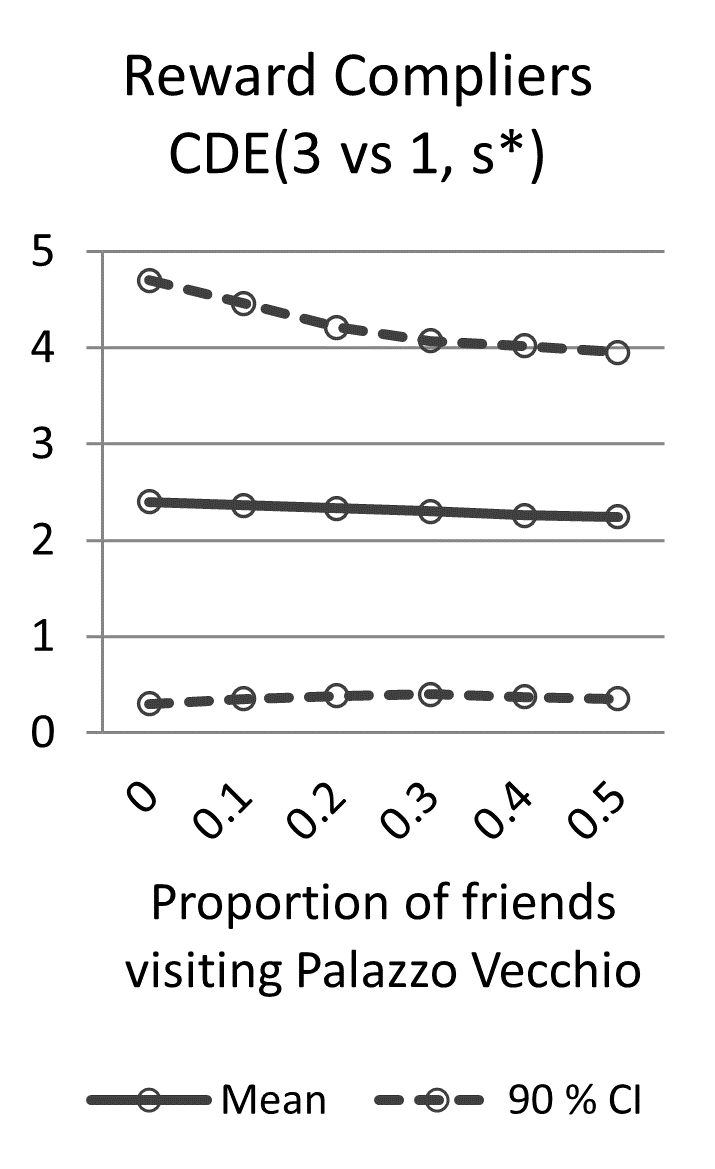}
	\label{fig:CDE}
\end{figure}

Principal controlled direct effects measure the causal effects of the encouragement in situations where the proportion of friends performing the visit to Palazzo Vecchio is arbitrarily set by the researcher to a fixed value $s$* for all units. In principle, any value $s^* \in [0,1]$ could be chosen. To avoid excessive extrapolation, we focus on a range of proportions that are not unreasonably distant from those actually supported by the data. %Indeed, it would make little sense to focus on situations where the proportion of friends visiting Palazzo Vecchio is so high that it cannot be  attained with the available encouragements. 
Figure \ref{fig:CDE} reports the CDEs evaluated at six different levels of $s^*$.

For Never Takers, it is worth showing only the CDE of the Presentation relative to the Flyer, $CDE_{000}(2 \text{ vs } 1,s^{*})$. As shown in Table \ref{tab:NDENIE}, the CDE of the Reward relative to the Presentation for Never Takers, $CDE_{000}(3 \text{ vs } 2,s^{*})$, is assumed to be zero because the Reward applies only to students who actually visit Palazzo Vecchio, and disappears in the follow-up period. As a consequence, $CDE_{000}(3 \text{ vs } 1,s^{*})=CDE_{000}(2 \text{ vs } 1,s^{*})$. The posterior means of the CDEs of the Presentation relative to the Flyer for Never Takers  show some evidence that there exists a positive direct effect of Presentation versus Flyer, which is approximately constant, equal to 1.65 additional museum visits, irrespective of the selected proportion of friends visiting Palazzo Vecchio. The 90\% posterior predictive  intervals cover zero, even if zero is very close to the lower bounds of the intervals, especially for proportion of  friends visiting Palazzo Vecchio lower  than (or equal to) 0.3.

For Reward Compliers, it makes sense to consider CDEs relative to any encouragement contrast (see previous discussion and Table \ref{tab:NDENIE}). $CDE_{001}(2 \text{ vs } 1,s^{*})$ can be interpreted as a pure encouragement effect, $CDE_{001}(3 \text{ vs } 2,s^{*})$ as the effect of the visit experience at Palazzo Vecchio, and $CDE_{001}(3 \text{ vs } 1,s^{*})$ as the sum of the previous two quantities. The CDEs of the Presentation relative to the Flyer for Reward Compliers are to raise the number of subsequent visits by more than 4 points, regardless of the fixed proportion of friends visiting Palazzo Vecchio. Therefore, the presentation has the effect of raising the intrinsic motivation of Reward Compliers to visit museums in the follow-up period. The posterior mean of the CDEs of Reward versus Presentation for Reward Compliers is negative but statistically negligible; the 90\% posterior credible intervals cover zero and are quite large. Therefore, there is no evidence that the Palazzo Vecchio experience may prompt future museum visits for Reward Compliers.

\begin{figure}[ht]
	\centering
	\caption{Estimated CSEs for Never Takers and Reward Compliers}
	\label{fig:CSE}
	\includegraphics[width=\textwidth]{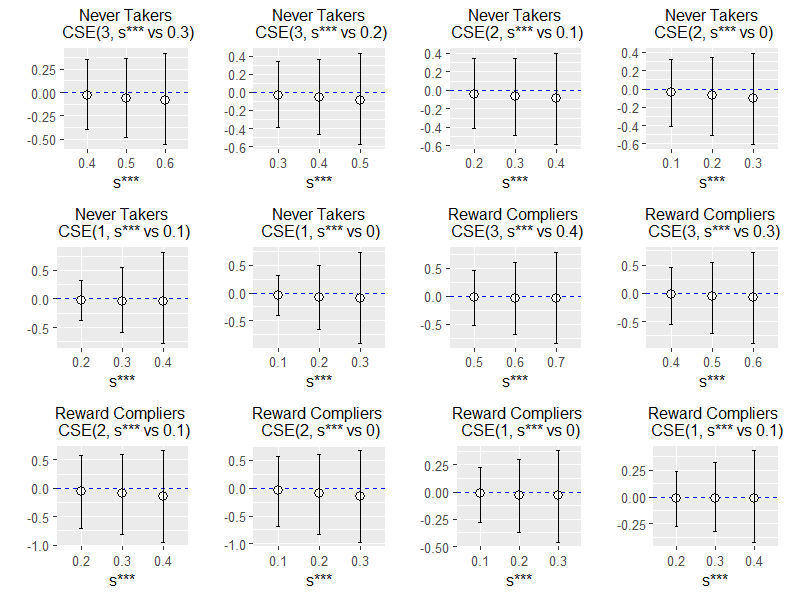}
	\begin{flushleft}
		\textit{Note}. The blue dashed line indicates zero.
	\end{flushleft}
\end{figure}

Finally, Figure \ref{fig:CSE} shows the means and the 90\% credible intervals of the posterior distributions of CSEs, originating from some discretionary change in the proportion of friends visiting Palazzo Vecchio while the encouragement is kept at a fixed level. 
In particular, the CSEs are evaluated at some alternative variations of such proportion from/to an arbitrary level that resembles the average proportion of friends visiting Palazzo Vecchio under the encouragement considered. This allows to evaluate how spillovers change when the proportion of friends visiting Palazzo Vecchio is higher or lower with respect to the latter. 

\noindent The spillover generated by any of these changes on the future museum attendance of both Never Takers and Reward Compliers are small and statistically negligible: the posterior means are close to zero and the 90\% posterior credible intervals cover zero and are approximately symmetric around zero. Therefore, no matter how much an individual is exposed to friends visiting Palazzo Vecchio, spillovers are hardly relevant in the decision to attend museums in the follow-up period.

\section{Concluding Remarks}

%This article shows how information on social networks may be used to investigate spillover effects within cluster encouragement designs, where individual non-compliance is usually an issue.  To this end, we propose  an original methodological interplay between principal stratification and mediation analysis, where a synthesis of the network information is used as a mediator which may generate spillover effects.

In this article  we show how information on social networks can be used to investigate spillover effects within cluster encouragement designs with individual non-compliance.  
We propose  an original methodological interplay between principal stratification and mediation analysis, where a synthesis of the network information is used as a mediator which may generate spillover effects. We formally define principal natural direct and indirect effects and principal controlled direct and spillover effects, and explain how these quantities can be interpreted to get information on different causal pathways, including spillover. 
We revisit results from a small field experiment, based on a clustered encouragement design, conducted in Florence (Italy) to study how appropriate personal incentives may lead students to visit museums in their free time. Possible causal pathways can either originate from such incentives directly affecting the student’s outcome, or follow an indirect trail where the student’s outcome is finally affected by the spillovers received from other students involved in the experiment. We use a Bayesian approach to inference under latent sequential ignorability assumptions. Previous causal studies of these experimental data \citep{Lattarulo:2017, Forastiere:2019} have not exploited the available information on friendship networks within each cluster of students, which we use here to disentangle and estimate spillover effects, and have thus settled for causal effects where spillovers are usually blended with other sources of change in the outcome.
Since our application is based on a small field experiment, the conclusions we reach would perhaps require corroboration in similar studies conducted on a larger scale. 
However, the results of our analysis suggest that spillovers from friends did not lead to significant changes in the students' museum attendance, the Presentation encouragement  was effective in motivating students to devote some free time to museum attendance, the visit experience that students were offered at Palazzo Vecchio  was not  effective, at least for the subpopulation of Reward Compliers.

%while some of the provided incentives did. In particular, the Presentation encouragement proved to be effective in motivating students to devote some free time to museum attendance. On the contrary, the visit experience that students were offered at Palazzo Vecchio does not seem to be effective, at least for the subpopulation of Reward Compliers, who would visit Palazzo Vecchio only if assigned to the Reward encouragement.

\section*{Acknowledgements}
The authors thank schools, teachers and students who participated in the study, as well as the
educational experts of the Mus.e Association for having performed the classroom presentations. The authors would also like to thank Laura Forastiere for her valuable comments.

\end{document}